\documentclass[journal,twoside,web]{ieeecolor}
\usepackage{generic}

\let\labelindent\relax
\usepackage{amsmath,amssymb,amsfonts,amsthm}
\usepackage[font=small]{caption}
\usepackage[font=small]{subcaption}
\usepackage{algorithmic}
\usepackage{graphicx}
\usepackage{textcomp}
\usepackage[ruled]{algorithm2e}
\SetKw{Break}{break}
\SetKw{Continue}{continue}
\usepackage{cuted}
\usepackage{enumitem}
\setitemize{noitemsep,topsep=0pt,parsep=0pt,partopsep=0pt}
\usepackage{dsfont}
\usepackage{orcidlink}
\usepackage[backend=biber,style=numeric-comp,sorting=none,natbib=true,doi=false,isbn=false,url=false,eprint=false,giveninits=true]{biblatex}
\usepackage{tikz}
\usetikzlibrary{automata, positioning}
\usepackage{pgfplots}
\usepackage{pgfplotstable} 

\usepgfplotslibrary{fillbetween} 
\pgfplotsset{compat=1.17}
\usepgfplotslibrary{colormaps}

\definecolor{mycolorZ0}{named}{red}
\definecolor{mycolorZ1}{named}{green}
\definecolor{mycolorZ2}{named}{blue}
\definecolor{mycolorZ3}{named}{orange}
\definecolor{mycolorZ4}{named}{cyan}
\definecolor{mycolorZ5}{named}{magenta}

\definecolor{mycolorA0}{RGB}{0,0,255}
\definecolor{mycolorA3}{RGB}{0,200,0}
\definecolor{mycolorA2}{RGB}{0,100,0}
\definecolor{mycolorA1}{RGB}{0,200,200}
\definecolor{mycolorA4}{RGB}{50,50,50}

\newtheorem{remark}{Remark}
\newtheorem{lemma}{Lemma}

\newtheorem{theorem}{Theorem}
\newtheorem{corollary}{Corollary}
\newtheorem{proposition}{Proposition}
\newtheorem{assumption}{Assumption}

\renewcommand{\baselinestretch}{1}

\DeclareMathOperator*{\argmax}{argmax} \allowdisplaybreaks
  
\usepackage{ifthen}
\newboolean{showcomments}
\setboolean{showcomments}{true}
\usepackage{todonotes}

\newtheorem{example}{Example}

\newboolean{arxiv}
\setboolean{arxiv}{false}

\definecolor{bleudefrance}{rgb}{0.19, 0.55, 0.91}
\definecolor{ao(english)}{rgb}{0.0, 0.5, 0.0}

\newcommand{\addcite}[0]{\ifthenelse{\boolean{showcomments}}
{\textcolor{purple}{(add cite(s)) }}{}}%

\newcommand{\addcites}[0]{\ifthenelse{\boolean{showcomments}}
{\textcolor{purple}{(add cite(s)) }}{}}%

\newcommand{\leopoldo}[1]{  \ifthenelse{\boolean{showcomments}}
{\todo[inline,color=bleudefrance]{Leopoldo: #1}}{}}

\newcommand{\juan}[1]{  \ifthenelse{\boolean{showcomments}}
{\todo[inline,color=pink]{Juan: #1}}{}}

\newcommand{\santiago}[1]{  \ifthenelse{\boolean{showcomments}}
{\todo[inline,color=orange]{Santiago: #1}}{}}

\newcommand{\miguel}[1]{  \ifthenelse{\boolean{showcomments}}
{\todo[inline,color=purple!60!white]{Miguel: #1}}{}}

\newcommand{\sean}[1]{  \ifthenelse{\boolean{showcomments}}
{\todo[inline,color=green!60!white]{Sean: #1}}{}}

\newboolean{showedits}
\setboolean{showedits}{true}
\usepackage[markup=underlined]{changes}
\definechangesauthor[color=bleudefrance]{EM}
\newcommand{\aem}[1]{
\ifthenelse{\boolean{showedits}}
{\added[id=EM]{#1}}
{\!#1\hspace{-4.75pt}}
}
\newcommand{\repem}[2]{
\ifthenelse{\boolean{showedits}}
{\replaced[id=EM]{#1}{#2}}
{\!#1\hspace{-4.75pt}}
}
\newcommand{\dem}[1]{
\ifthenelse{\boolean{showedits}}
{\deleted[id=EM]{#1}}
{}
}

\def\BibTeX{{\rm B\kern-.05em{\sc i\kern-.025em b}\kern-.08em
    T\kern-.1667em\lower.7ex\hbox{E}\kern-.125emX}}
\begin{document}
{\setbox0=\hbox{\includegraphics{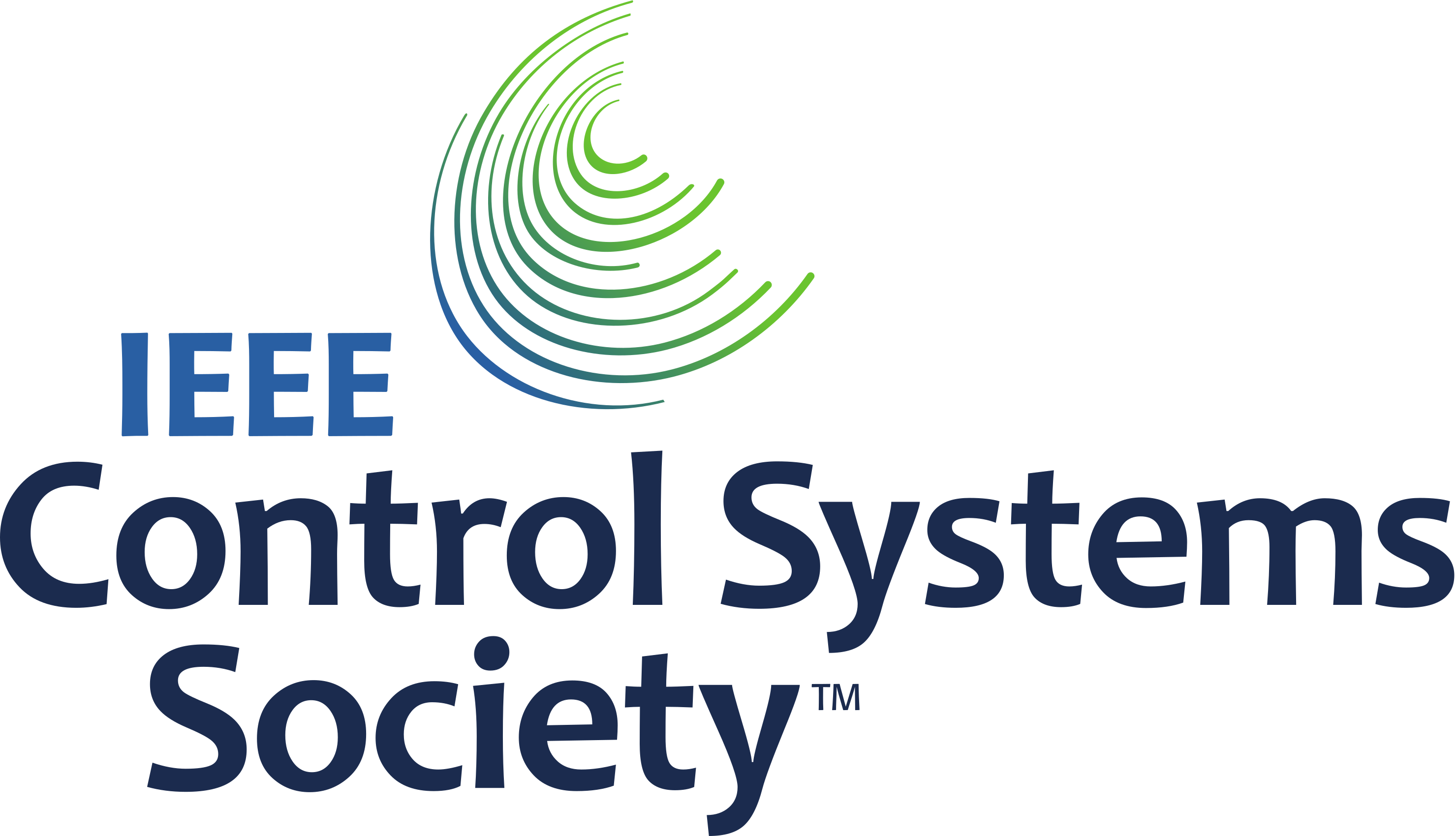}}}
\title{
Cooperative Multi-Agent Assignment over Stochastic Graphs via Constrained Reinforcement Learning
}

%

\author{Leopoldo Agorio, Sean Van Alen, 
         Santiago Paternain,
        Miguel Calvo-Fullana, and 
        Juan Andres Bazerque      
\thanks{Leopoldo Agorio, is with the Dept. of Electrical Engineering, Universidad de la Republica, Montevideo, 11300, Uruguay.   (email: lagorio@fing.edu.uy). \\
Sean Van Allen, is with the Dept. of Electrical and Computer Engineering, Univ. of Pittsburgh, Pittsburgh, PA, USA  (email:  sgv13@pitt.edu). \\
Santiago Paternain is with the 
Dept. of Electrical Computer and Systems Engineering, Rensselaer Polytechnic Institute, Troy, NY, USA (e-mail: paters@rpi.edu).\\
Miguel Calvo-Fullana is with the Dept of Engineering,  Universitat Pompeu Fabra, Barcelona, Spain (email: miguel.calvo@upf.edu). \\
Juan Andres Bazerque is with the Department of Electrical and Computer Engineering, Carnegie Mellon
University, Pittsburgh, PA 15213 USA and with the Dept. of Electrical Engineering, Universidad de la Republica, Montevideo, 11300, Uruguay (e-mail: jbazerque@cmu.edu).   
}}
\maketitle

\begin{abstract}
  Constrained multi-agent reinforcement learning offers the framework to design scalable and almost surely feasible solutions for teams of agents operating in dynamic environments to carry out conflicting tasks. We address the challenges of multi-agent coordination through an unconventional formulation in which the dual variables are not driven to convergence but are free to cycle, enabling agents to adapt their policies dynamically based on real-time constraint satisfaction levels. The coordination occurs during deployment, and relies on a light single-bit communication protocol over a network with stochastic connectivity. Using this gossiped information, agents update local estimates of the dual variables. Furthermore, we modify the local dual dynamics by introducing a contraction factor, which lets us use finite communication buffers and keep the estimation error bounded. Under this model, we provide theoretical guarantees of almost sure feasibility and corroborate them with numerical experiments in which a team of robots successfully patrols multiple regions, communicating under a time-varying ad-hoc network. 
\end{abstract}

\section{Introduction}
Reinforcement Learning (RL) has emerged as a key paradigm in artificial intelligence, learning optimal behaviors through interaction with the environment. Many real-world scenarios involve multiple agents with their own objectives, observations, and decision-making processes, giving rise to Multi-Agent Reinforcement Learning (MARL) \cite{canese2021multi,zhang2021multi,yang2021believe}. Centralized solutions, i.e. deciding all actions from the full set of observations, struggle to scale with the number of agents, disregard individual agent autonomy, and often cannot handle decentralized information structures with limited or private observations \cite{zhang2021multi,busoniu2008comprehensive}. MARL instead leverages distributed approaches \cite{lin2021multi} where each agent decides based on its own observations, achieving strong results in multi-player games, cooperative tasks such as Hanabi \cite{bard2020hanabi}, emergent behavior in hide-and-seek environments \cite{baker2019emergent}, and real-world applications such as traffic light control \cite{wei2019colight}.

Control applications further showcase this adaptability, with advances in collision avoidance for autonomous vehicles \cite{kahn2017uncertainty,isele2018safe}, agile locomotion in legged robots \cite{yang2022safe,hwangbo2019learning}, and manipulation in industrial settings \cite{ibarz2021train,kober2013reinforcement}. Many of these applications require constrained formulations, where safety \cite{chen-ames2024probabilistic,chen-subramian2024probabilistic,garcia}, resource limits, or task priorities are encoded as mathematical inequalities, giving rise to Constrained Reinforcement Learning (CRL) \cite{liang2018accelerated,castellano2023learning}.

In this work, we focus on multi-agent assignment, a class of MARL problems in which agents are assigned to specific requirements, revising these assignments dynamically to rotate among requirements in a coordinated manner.
Constraints are essential to model conflicting requirements. Unconstrained assignment methods combining RL with game theory have been proposed, requiring minimal \cite{cui2020multi} or no coordination \cite{qin2021multi}, but these account only for local rewards and offer no guarantee that assignment specifications are met. In contrast, we propose a constrained optimization approach that guarantees satisfaction of global requirements involving the joint state of all agents.

We therefore address the assignment problem under Constrained Multi-Agent Reinforcement Learning (CMARL) \cite{gu2021multi,chen2020autonomous,liu2022distributed,ying2023scalable,zhao2023multi}, a literature that addresses safe coordination in autonomous fleets \cite{shalev2016safe}, fair resource allocation \cite{zimmer2021learning}, and robust cooperation under communication constraints \cite{xiao2023graph}. Constraints may be defined per agent \cite{gu2021multi}, or collectively, involving all agents in the feasibility of the same constraint \cite{chen2020autonomous,liu2022distributed}. Assignment problems entail multiple such collective constraints but do not admit the aggregate reward structure in \cite{chen2020autonomous,liu2022distributed}.

Whether applied to single or multiple agents, CRL methods often employ regularization to reconcile competing objectives, transforming constraints into penalty terms within an aggregate reward \cite{liang2018accelerated,liu2022deep}. For instance, a mobile robot navigating crowded environments might balance collision avoidance (hard) with energy efficiency (soft) by weighting their penalties (e.g., \cite{kahn2017uncertainty}). Selecting these weights is non-trivial: overly conservative penalties yield overly cautious behavior, while insufficient regularization risks constraint violations \cite{holder2024multi}. Bayesian methods \cite{hutter2002self} adapt weights probabilistically using prior knowledge of constraint criticality, while heuristic approaches such as reward shaping \cite{tessler2018reward}, though less rigorous, remain common in practice.

Lagrangian duality \cite{liang2018accelerated,paternain2019constrained,kahe2026distributed} offers a more principled approach, dynamically adjusting weights via gradient descent on the dual variables. Although the duality gap is guaranteed to be zero \cite{paternain2019constrained}, feasibility is not guaranteed when the primal maximizer is not unique~\cite{calvo2023state}, a case where both regularization- and duality-based methods relying on multiplier convergence struggle. Example \ref{example} in Section \ref{sec:problem_form} and the discussion in Section \ref{sec:offline_training} illustrate how regularization fails in assignment problems where the collective constraints are nonlinear combinations of individual ones.

State augmentation addresses these limitations by reinterpreting the constrained RL problem as an extended Markov Decision Process (MDP). Treating Lagrangian multipliers as part of the state space lets policies condition their actions on both environmental observations and current constraint satisfaction \cite{calvo2023state}, yielding a family of policies that alternate as the Lagrangian weights cycle to multiplex between them: as one policy attends to one requirement, the weights evolve within the augmented MDP to eventually force a shift toward a policy that tackles another unsatisfied requirement. Theoretical analyses show that such approaches preserve convergence guarantees under mild assumptions, effectively converting constrained optimization into an unconstrained problem over the augmented state-action space \cite{paternain2022safe}.

In our preliminary work \cite{agorio2024multi}, we leverage these extended MDPs so that multiple agents can solve assignment problems collaboratively via CMARL, following a centralized training-distributed execution framework \cite{ikeda2022centralized}. Since standard dual methods prove inadequate, the cycling multiplier trajectories in the augmented state drive agents to alternate between tasks. These multipliers are shared through a communication network, in line with decentralized networked RL~\cite{zhang2018fully}, enabling coordination without direct access to other agents' states and yielding a distributed protocol with theoretical feasibility and convergence guarantees. Building on this foundation, we introduce this work's key novelty: a gossiping framework \cite{aysal2009broadcast} to share multipliers across a stochastic communication network \cite{casteigts2012time,Nedic2017Achieving} that is inherently dynamic and may even be disconnected across time, reflecting ad-hoc conditions common in practice. To maintain feasibility under these conditions, we propose a contractive update for the augmented multipliers, which allows finite communication buffers while keeping the estimation error bounded, enhancing scalability and ensuring almost sure feasibility in stochastic environments. We prove almost sure feasibility up to an error margin that can be made arbitrarily small, via a proof technique that explicitly accounts for the stochasticity of the communication graph, and validate these contributions numerically: a team of five robots successfully learns to patrol six regions of interest, achieving coordination under realistic communication constraints.

The rest of this paper is organized as follows. Section \ref{sec:problem_form} introduces multi-agent assignment as a feasibility problem and presents the stochastic graph model governing agent communication, capturing the dynamic and potentially disconnected nature of typical networks. Section \ref{sec:offline_training} details the state augmentation approach that lets agents coordinate using dual variables shared through the network, and outlines the offline training and online execution algorithms. Section \ref{sec:feasibility} states and proves our main theoretical result: almost sure feasibility of the proposed algorithm under mild assumptions. Section \ref{sec_numerical} validates our approach numerically, with a team of five robots patrolling six regions of interest and demonstrating consistent feasibility across all runs. Finally, Section \ref{sec:conclusion} concludes, with supporting lemmas in the Appendix.

\section{Problem Formulation}
\label{sec:problem_form}

Consider $N$ agents interacting with an environment that is modeled as an MDP with transition probabilities $\mathbb{P}\left(S_{t+1}\mid S_{t},A_{t}\right)$. For each time index $t=0,1,2,\ldots$, random variables $S_t=(S_t^1,\ldots,S_t^N)$ and $A_t=(A_t^1,\ldots,A_t^N)$ collect the states $S_t^n\in \mathcal S$ and actions $A_t^n\in \mathcal A$ for all agents. The team receives a vector of rewards $r(S_t)\in \mathbb R^M$ associated with $M$ different tasks that must be collectively satisfied. We consider a family of assignment problems in which the agents must persistently cover $M$ regions of the state space denoted by $\mathcal{S}_m\subset\mathcal{S}$ with $m=1,\ldots,M$. Thus, each reward is given by
\begin{equation}\label{eqn_patrolling_reward}
    \left(r(S_t)\right)_m = \max_{n=1,\ldots,N} \mathds{1}[S_t^n\in \mathcal{S}_m],
\end{equation}
where  $\left(r\right)_m$ represents the $m$-th element of vector $r$,  with the index function defined as
    $\mathds{1}[S_t^n\in \mathcal{S}_m]=1 $ if $  S_t^n\in \mathcal{S}_m$ and zero otherwise. 
Each region $\mathcal S_m$ must be covered by the agents a fraction of the time. Hence, we set the requirements as a vector $c\in[0,1]^M$ of $M$ thresholds on the average rewards, and formulate a Markov decision problem with constraints
\begin{equation}\label{eqn_ideal_V}
    V(\pi):= \lim_{T\to \infty}\frac{1}{T} \mathbb E_{S_t, A_t\sim \pi} \left [ \sum_{t=0}^{T-1} r(S_t) \right ] \geq c,
    \end{equation}
with the inequality being component-wise. The expectation is taken over the transition probabilities and the probability distribution defining the control policy $\pi(A_t|S_t)$  of the actions of all agents given their states.

A task reward can be incorporated to establish an optimization objective. This can be useful to set a secondary goal, such as minimizing energy consumption in a multi-robot application like the one presented in Example \ref{example} below. However, in the following, we will set the objective to zero and focus on our primary goal of satisfying the specifications.

Hence, the feasibility problem we aim to solve is
\begin{align}\label{eqn_crl}
P^\star = &\; \max_{{\pi}} \ 0 \\*
\textrm{s. to} &\; \lim_{T\to \infty}\frac{1}{T}  \mathbb E_{S_t, A_t\sim \pi} \left [ \sum_{t=0}^{T-1} r(S_t) \right ] \geq c.\notag 
\end{align}
To illustrate the practical use of this problem formulation, we include the following example, which we will explore further in the experiments of Section \ref{sec_numerical}.

\begin{figure}[t]
    \centering
	\includegraphics[scale=1]{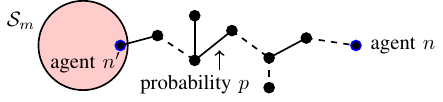}    
    \caption{\label{fig:gossip_stochastic}Agent $n$ must receive the information that agent $n^\prime$ is in zone $\mathcal S_m$ across the communication graph. Agents \( n \in \mathcal{V} = \{1, \dots, N\} \) are nodes in the stochastic graph \( \mathcal{G}(\mathcal{V}, \mathcal{E}, p) \), where solid and dashed lines represent the presence or absence of an edge, respectively, at a particular time. If an edge is present, the two nodes connected to the edge can exchange information. Otherwise, the two neighboring agents must wait for a future time in which the edge is present. }
\end{figure}

\begin{example}\label{example}
Consider a team of $N$ robots that is required to monitor $M>N$ areas,  visiting areas $\mathcal S_1,\ldots,\mathcal S_M$ during fractions of time given by  $c\in [0,1]^M$. These conflicting requirements may not be achievable by a single robot, particularly when the aggregate thresholds are greater than one, i.e., $\left\|c\right\|_1>1$, and not even by multiple robots acting independently following the same optimal single-agent policy. Coordination is thus essential, for which robots can form an ad-hoc communication network. 
\end{example}
We aim to obtain a feasible policy for (3) that can be deployed in a coordinated distributed fashion, in a scenario in which agents do not have access to the whole system state but can only observe their own  $S_t^n$. Once deployed, agents will not share their locations. Instead, we will develop a light gossip protocol for agents to exchange \emph{single-bit messages} with their neighbors (Section \ref{sec_online_execution}) over an ad-hoc communication network adhering to the stochastic graph model in \cite{Nedic2017Achieving}.  This graph model considers a time domain $\mathcal T \subset \mathbb N$ and a set of nodes $\mathcal V=\{1,\ldots,N\}$ representing the agents. Then, $\mathcal G = \mathcal G(\mathcal V,\mathcal E,\mathcal T, w_G)$ defines a stochastic graph where edges are sampled from a static underlying graph $G = (\mathcal V, \mathcal E)$. This \textit{footprint} graph $G$ indicates pairs of agents that could have connectivity at some point in time. In this model,  $G$ is assumed to be connected, but it is noteworthy that this does not necessarily imply that $\mathcal G$ is connected. Indeed, $\mathcal G$ could be disconnected at all times. Accordingly, the \textit{presence} function $w_G: \mathcal E \times \mathcal T \to \{0,1\}$ indicates if an edge is present at a given time.
We will consider a probabilistic model in which the presence of an edge 
$e^{(n,n^\prime)} \in\mathcal{E}$ 
between two agents    $n,n^\prime \in \mathcal{V}$  at time $t$ is modeled as a Bernoulli random variable $w_G$ with probability $p$, independent across edges and time
\begin{align}
w_G\left(n,n^\prime,t\right)& \text{ are i.i.d. }  \text{Bernoulli}(p).\label{eqn_graph_model}
\end{align} 
   Figure \ref{fig:gossip_stochastic} illustrates the communication graph for the monitoring problem in Example \ref{example}.

\section{Offline Training and Online Execution}
\label{sec:offline_training}

This section presents two algorithms for training the policy and carrying out the tasks. Combined, they solve the CRL problem \eqref{eqn_crl} in the dual domain~\cite{paternain2019constrained}.

Define the Lagrangian associated with problem \eqref{eqn_crl}
\begin{equation}\label{eqn_lagrangian}
\mathcal{L} (\pi,\lambda) = \lim_{T\to \infty}\frac{1}{T} \mathbb E_{S_t, A_t\sim \pi} \left [ \sum_{t=0}^{T-1} \lambda^\top\left(r(S_t)-c\right)\right ],
\end{equation}
where  $\lambda \in\mathbb{R}_+^M$ denote the dual variables. 

To obtain the dual, we maximize the Lagrangian in \eqref{eqn_lagrangian}. 
However, we must divert from \emph{standard} primal-dual algorithms~\cite{arrow1958studies} that aim for a dual-optimal vector of multipliers.
That is because problem \eqref{eqn_crl} is such that primal-dual pairs are degenerate, as the Lagrangian does not have a unique maximizer when weighted by the dual optimal multipliers. To simplify the explanation, consider an abridged version of Example 1 in which a single agent needs to cover two areas. For a set of dual variables to be optimal, it must assign the same weight to each area. Otherwise, the agent would focus solely on maximizing the reward associated with the area with the higher multiplier, thus visiting only that region. But since the other area is unattended, this policy is infeasible. Even if we consider a dual optimal variable\textemdash one that assigns the same reward to the two areas\textemdash multiple primal solutions are possible, and a feasible one is not guaranteed. The primal optimal policy may direct the agent to visit both areas as desired, or, with the same reward, could focus on a single region. This policy would maximize the Lagrangian but would not satisfy the conflicting constraints, so that it would still be infeasible for the constrained problem \eqref{eqn_crl}. Since the primal maximizer is not unique, and the optimal multipliers do not provide useful information to choose between these two solutions, we cannot guarantee that the feasible one will be found. The same fundamental problem persists if we consider a larger case with more zones than agents.



This limitation of primal-dual methods is not specific to our formulation but was demonstrated in \cite{calvo2023state}  and appears even in convex optimization when the Lagrangian is not strictly convex, in which case iterates need to be averaged to achieve convergence~\cite{nedic2009subgradient}. However, we cannot rely on averaging for our assignment problem \eqref{eqn_crl}  since the argument for dual optimality of the averaged multipliers relies on the convexity of the primal.

In our constrained RL scenario, we aim to solve problem \eqref{eqn_crl} by an alternative two-step process. First, in an offline training stage, we obtain a policy that optimizes the Lagrangian in \eqref{eqn_lagrangian} for all $\lambda$. In the deployment stage, we apply a stochastic dual gradient descent iteration, which causes cycles in $\lambda$, and let the agents choose a different policy  per rollout, thus preventing an infeasible policy resulting from convergence of $\lambda$ to the set of optimal multipliers. As multipliers cycle, the rewards in \eqref{eqn_lagrangian} are reweighted, the highest entries of $\lambda$ alternate, and thus the policy maximizing \eqref{eqn_lagrangian} gives priority to constraints that have not being attended in previous cycles.     

{Before describing these two stages in detail,  we introduce the following two practical considerations into our design. 
First, to prevent agents from requiring a full observation of the global $S_t$ and  shield our policy from the  exponential growth  of the state space, we introduce the following structure
$\pi(A_t|S_t,\lambda^1,\ldots,\lambda^N) = \prod_{n=1}^N \pi^n(A_t^n| S_t^n,\lambda^n)$, in which each agent decides its action taking into account its state $S_t^n$  and local copies $\lambda^n$ of $\lambda$ only. 

Note that even if the policies are local,  agents are still coupled by their dynamics and rewards. Therefore, the agents need to optimize the Lagrangian \eqref{eqn_lagrangian}  jointly, i.e., 
\begin{align}\label{eqn_optimal_set}
&\hspace{-8pt}\pi^\star[\lambda,\ldots,\lambda]\hspace{-2pt}=\hspace{-2pt}\argmax_{\pi[\lambda,\ldots,\lambda]} \lim_{T\to \infty}\hspace{-1pt}\frac{1}{T}\hspace{-1pt}\mathbb E_{S_t, A_t\sim \pi}\hspace{-3pt}\left [\hspace{-1pt}\sum_{t=0}^{T-1}\hspace{-3pt}r_{\lambda}(S_t)\hspace{-1pt}\right]\hspace{-1pt},
\end{align}
in which we reduced notation by optimizing over the global policy $\pi[\lambda,\ldots,\lambda]$  defined as 
\begin{align}\label{eqn_ideal_separated_policy}\pi[\lambda,\ldots,\lambda](A_t|S_t) &:=\pi(A_t|S_t,\lambda,\ldots,\lambda)\\&=\prod_{n=1}^N \pi^n(A_t^n|S_t^n,\lambda),\label{eq:factorized_policy}
\end{align}
and simplified \eqref{eqn_lagrangian} by defining the following weighted reward 
\begin{equation}\label{eqn_reward_lambda}
    r_\lambda(S_t) :=  \lambda^\top\left(r(S_t)-c\right).
\end{equation}
The factorization in \eqref{eq:factorized_policy} introduces a  simplifying policy design in which each agent only requires to observe its own state, recalling that  state and action of a single agent at time $t$ is described by the superscript $n$ as in $S_t^n,A_t^n$, as opposed to $S_t,A_t$ which refers to the entire team.
Furthermore, we parameterize each agent's policy by a vector $\theta^n \in \mathbb{R}^{p_n}$. A judicious parameterization will allow us to deal with the augmented state-space $\mathcal{S}\times {\mathbb{R}^M_+}$,  containing elements $(S_t,\lambda)$, where at least the multipliers are continuous variables. Hence, each agent's action is drawn from a distribution~$\pi_{\theta^n}(A_t^n \mid S_t^n, \lambda)$, so that the joint probability distribution with parameters $\theta=(\theta^1,\ldots,\theta^N)$ is 
\begin{align}\label{eqn_separate_policy}
    \pi_\theta[\lambda,\ldots,\lambda](A_t\mid S_t)&:=\pi_\theta(A_t\mid S_t,\lambda,\ldots,\lambda)\\&=\prod_{n=1}^N\pi_{\theta^n}(A_t^n\mid S_t^n,\lambda).
\end{align}  
Next, we describe how to train for these parameters $\theta$.

\subsection{Offline Training}\label{ssec_offline_training}
The policy structure in \eqref{eqn_separate_policy} induces a simplified form of the policy gradient \cite{sutton2000policy} that we will use for training. 
Before presenting this result in Proposition \ref{prop_gradients}, we define the following key concepts.
The occupancy measure for the joint state and actions of all agents is given by
%
$  \rho_{\theta,\lambda}(s,a) = \lim_{t\to \infty}\frac{1}{t}\sum_{\tau=0}^t \mathbb{P}(S_\tau=s)\pi(a|s).$
Likewise,   the state-action value function is defined by $Q_{\pi_\theta}(s,a,\lambda)\text{=} \sum_{t=0}^{\infty} \mathbb{E}_{\pi_\theta}\left[r_\lambda(S_t)\text{-}\mathcal L(\pi_\theta,\lambda)| S_0\text{=}s,A_0\text{=}a\right],$ 
%
where $r_\lambda(S_t)$ and $\mathcal{L}(\pi_\theta,\lambda)$ are the functions defined in \eqref{eqn_reward_lambda} and \eqref{eqn_lagrangian}, respectively. 
We further define the state-action value function for agent $n$ as
\begin{equation}\label{eqn_q_others}
    Q^{n}_{\pi_\theta}(s^n,a^n,\lambda) = \mathbb{E}\left[Q_{\pi_\theta}(S_t,A_t,\lambda)\mid S_t^n=s^n, A_t^n=a^n \right]. 
\end{equation}
with expectation probability $\rho_{\theta,\lambda}(S_t,A_t)$ over $(S_t,A_t)$.
From the perspective of agent $n$, the local Q-function \eqref{eqn_q_others} is the average of the expected return, assuming that all other agents follow the policy $\pi_\theta[\lambda,\ldots,\lambda]$. Therefore,  $Q^{n}_{\pi_\theta}(S_t,A_t,\lambda)$ can be estimated per agent by averaging the global rewards that result from their local states and actions, or modeled as a parametric function of these local entries. The following proposition formalizes these insights, providing a specific form for the policy gradient Theorem under the separable policy design.

\begin{proposition}\label{prop_gradients}
Assume that the policy of each agent is parameterized by a vector $\theta^n\in\mathbb{R}^n$ and that $A^n\sim\pi_{\theta^n}(\cdot| S^n,\lambda)$, where  $S^n$ represents the local state of agent $n$. Let $\mathcal{L}(\pi_\theta,\lambda)$ and $Q^{n}_{\pi_\theta}(S^n,A^n,\lambda)$ be the functions defined in \eqref{eqn_lagrangian} and \eqref{eqn_q_others} respectively. Then, it follows that
%
  $  \nabla_{\theta^n} \mathcal{L(\pi_\theta,\lambda)} = \mathbb{E}\left[\nabla_{\theta^n }\log \pi_{\theta^n}(A^n|S^n,\lambda)Q^{n}_{\pi_\theta}(S^n,A^n, \lambda)\right]$
where the expectation is taken with ${(S^n,A^n) \sim \rho_{\theta,\lambda} }.$

\end{proposition}
\begin{proof}See Appendix \ref{app:proof_prop_gradients}.
\end{proof}

Proposition \ref{prop_gradients} tells us that agent $n$ can compute \emph{locally} its part of the gradient $\nabla_{\theta^n} \mathcal{L(\pi_\theta,\lambda)}$, the part needed to update its own local parameters $\theta^n$, by using its local policy $\pi_{\theta^n}(A_t|S_t,\lambda)$, the multiplier $\lambda$, and the local Q-function in \eqref{eqn_q_others}  averaging over the state and actions of the other agents.
Hence, we leverage this local property of the gradients to optimize a  parametric version of \eqref{eqn_optimal_set}
\begin{align}\label{eqn_optimal_trained_policy}
&\pi_{\theta^\star}[\lambda,\ldots,\lambda]=\argmax_{\pi_{\theta}[\lambda,\ldots,\lambda]}\mathbb E_{S_t, A_t\sim \pi_\theta}\left[\frac{1}{T_0}\sum_{t=0}^{T_0-1}r_{\lambda}(S_t,A_t)\right],
\end{align}
where we used a slight abuse of notation to write a policy $\pi_\theta[\lambda,\ldots,\lambda]$ complying with \eqref{eqn_separate_policy} as the optimization variable instead of $\theta$.     Notice that, in addition to the parametric expansion of the policies, we introduced a second variation by removing the limit and adhering to a finite time horizon. This modification is key for the practical implementation of the training algorithm, which is detailed in Algorithm \ref{alg:offline}. {Let us remark that the policy parameters that result from   \eqref{eqn_optimal_trained_policy} do not depend on $\lambda$. The augmented state $(S_t^n,\lambda)$  is the input of each local policy $\pi_{\theta^n}$ with parameters $\theta^n$ as in \eqref{eqn_separate_policy}.
}

 Next, we formalize our assumptions on the errors induced by the modifications above (parameterization and finite horizon). For this purpose, we  define the truncated value $V_{T_0}(\pi)$ which becomes  
\begin{equation}\label{eq:non-idealities-VP}
V_P:=V_{T_0}(\pi_\theta[\lambda,\ldots,\lambda]):=\mathbb E_{\pi_\theta}\left[\frac{1}{T_0} \sum_{t=0}^{ T_0-1} r(S_t)\ \right],
\end{equation}
under policy parameterization, so that $V_P$ corresponds to the finite-horizon expected cumulative reward attained by the team when running the structured policy \eqref{eqn_separate_policy} for a particular value of $\lambda$ common to all agents.
\begin{assumption}\label{assumption_representation}(Approximation error in training).
The policy parameterization is dense enough, and the horizon is sufficiently large, i.e.,
 For any $\epsilon>0$ there exists $\beta>0$ and $T_0>0$, such that for each  policy $\pi[\lambda,\ldots,\lambda]$ in \eqref{eqn_ideal_separated_policy} there exists a set of parameters $\theta$ and associated  policy $\pi_\theta[\lambda,\ldots,\lambda]$ in \eqref{eqn_separate_policy} such that the respective values $V=V(\pi[\lambda,\ldots,\lambda])$ and $V_P=V_{T_0}(\pi_\theta[\lambda,\ldots,\lambda])$ defined in  \eqref{eqn_ideal_V}  and  \eqref{eq:non-idealities-VP}, satisfy
\begin{align}\label{eqn_representation_error}
\lambda^\top(V-V_P)&\leq \beta \|\lambda\|+\epsilon.
\end{align}
\end{assumption}
This assumption is tantamount to a universal approximation property, which is standard, for instance, if the policies are parameterized via neural networks, whose outputs are the probabilities assigned to the set of actions. 
Specifically, if the neural network parametrization is expressive enough, then the corresponding value $V_P$ approximates the full $V$  and $\beta$ becomes small.
The term $\epsilon$ in \eqref{eqn_representation_error}  is introduced to bound the truncation error. This bound is not an assumption actually, but results from the rewards in \eqref{eqn_patrolling_reward} being bounded.  In addition, $\epsilon$ can accommodate the limited accuracy in solving \eqref{eqn_optimal_trained_policy} that results if running a finite number of optimization iterations and the error in the gradient of the Lagrangian computed over a finite horizon.
\begin{algorithm}[t]
\caption{Offline training loop for agent $n$}
\label{alg:offline}
\begin{algorithmic}[1]
\FOR{$k=0,1,\ldots$}
  \STATE Sample $\lambda\in \Lambda\subset \mathbb  R_+^M$
    \STATE Sample $S_0^n \sim \mathcal S$
    \FOR{$t=0,\ldots,T_0$}
      \STATE Sample $A_t^n \sim \pi_{\theta^n}(A_t^n \mid S_t^n, \lambda)$
      \STATE Update $S_{t+1}$
      \STATE Compute $r(S_{t+1})$
    \ENDFOR
    \STATE Compute $\nabla_{\theta^n} \mathcal{L(\pi_\theta,\lambda)}$ as in Proposition \ref{prop_gradients}
    \STATE Update $\theta^n\leftarrow \theta^n+\eta  \nabla_{\theta^n} \mathcal{L(\pi_\theta,\lambda)}$
\ENDFOR
\end{algorithmic}
\end{algorithm}
%

Training for \eqref{eqn_optimal_trained_policy} requires that all agents interact to learn from rewards \eqref{eqn_reward_lambda} that are jointly activated via \eqref{eqn_patrolling_reward}. This training phase is designed to run offline, with multiple agents and a common $\lambda$ drawn randomly at the episode start and fixed until it ends.  Once each policy $\pi_{\theta^n}(A_t^n|S_t^n,\lambda)$ in \eqref{eqn_optimal_trained_policy} has been trained, its online execution depends only on the local state $S_t^n$ of the agent and a common $\lambda$ which varies as agents enter and exit the zones. Thus, agents must engage in online coordination to concur on 
the trajectory of $\lambda$, described next.

\subsection{Online Execution}
\label{sec_online_execution}

In deployment, each agent will use the vector $\lambda$ as a warning signal that a particular constraint needs attention. Even if we do not aim for the dual optimal multipliers, we still derive our intuition from dual theory, considering in particular the standard gradient update    $\lambda_{k+1} = \left[\lambda_{k}-\eta\left(V\left(\pi\left[\lambda_{k},\ldots,\lambda_k\right]\right)-c\right)\right]_+,$ where $[\cdot]_+$ denotes the projection on the non-negative orthant.    Following this recursion, a multiplier in $\lambda_k$ associated with a constraint that is not being properly addressed will increase, and one whose value surpasses the required threshold will tend to zero.       \

For a data-driven implementation, we run a stochastic version of this dual update  \cite{dvoretsky1955stochastic}, dropping the expectation and replacing it with reward samples. As we did for training, we also set a finite horizon $T_0$.   

Hence, we use the following stochastic dual update 
\begin{align}\label{eqn_contractive_big_brother}
    \hspace{-6pt}\lambda_{k+1}\hspace{-2pt}=\hspace{-2pt}\left[(1-\alpha)\lambda_k\hspace{-1pt}+\hspace{-1pt}\frac{\eta}{T_0}\hspace{-2pt}\sum_{\tau=k T_0}^{(k+1)T_0-1}\hspace{-6pt}\left(c-r(S_\tau) \right)\right]_{+}\hspace{-2pt},
\end{align}
introducing a contractive factor $(1-\alpha)\in(0,1)$, which is necessary to ensure that the errors induced by the communication graph do not grow unbounded, as we discuss on the remark at the end of this section.

The stochastic update in \eqref{eqn_contractive_big_brother} can be incorporated into the MDP modeling $S_t$ to form an augmented dynamical system with state  $(S_t,\lambda_k)$ in two timescales, with $t$ representing the time step and $k$ indexing the rollout, which spans $T_0$ time steps from $kT_0$ to $(k+1)T_0-1$. 

During the execution phase, the agents cannot observe global rewards, which are available locally only to the agents whose states $S_t^n$ are in regions $\mathcal S_m$. Hence, the update in \eqref{eqn_contractive_big_brother} is not realizable. Instead, agents exchange the reward information across the communication network to implement \eqref{eqn_contractive_big_brother} in a distributed manner. To devise such a distributed procedure, define $R_{\tau,t}^n\in\mathbb{R}^M$ as the estimate that the agent $n$ has at time $t$ about the actual value of the rewards $r(S_\tau)$ at a previous time $\tau\leq t$. The communication timeline is shown in Figure \ref{fig:timeline}.
\begin{figure}[t]
    \centering
	\includegraphics[scale=1]{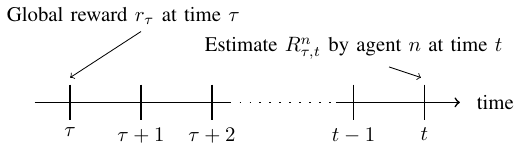}        
    \caption{\emph{Gossip} timeline: Each agent aims to estimate the global vector of rewards $r(S_{\tau})$. After $t-\tau$ time steps, the local estimation of $r(S_{\tau})$ obtained by agent $n$ is $R_{\tau,t}^n$.}
    \label{fig:timeline} 
\end{figure}
Consider an agent $n^\prime$ whose state $S_\tau^{n^\prime}$ is in region $\mathcal S_m$  at time $\tau$. That is $S_\tau^{n^\prime}\in \mathcal S_m$ as shown in Figure~\ref{fig:gossip_stochastic} for the case of our monitoring Example \ref{example}. Then agent $n^\prime$ knows first-hand that the reward for the region $S_m$ is being attended at this specific time $\tau$, and thus the global reward $(r(S_{\tau}))_m = 1$ is being attained. Hence, it sets its own local estimate of the reward to one,  i.e., $(R_{\tau,t}^{n^\prime})_m = (r(S_{\tau}))_m = 1 \ \textrm{for all}\ t\geq \tau$. Then, it passes this information to its neighbors, who will update their local estimates of the reward $(r(S_{\tau}))_m$ accordingly. Specifically, each agent will receive the local estimates of $r(S_{\tau})$ from its neighbors and set its local estimate to $1$ if any of its neighbors have a local estimate of $1$ for the same region. That is equivalent to using the maximum of the local estimates of its neighbors, including itself. In general, the update rule for the local copies of the rewards is given by
\begin{align}\label{eqn_gossip_r}
    	\left(R_{\tau,t}^n\right)_m&=\begin{cases}
    	\max_{n^\prime\in\mathcal N_n\cup \{n\}} \left(R_{\tau,t-1}^{n^\prime}\right)_m,& t>\tau\\
    	\mathds 1\{S_{t}^n\in\mathcal S_m\},& t=\tau,
    	\end{cases}
\end{align}
where $\mathcal N_n$ is the (stochastic, time-varying) neighborhood of agent $n$ at time $t$, and $\mathds 1\{S_{t}^n\in\mathcal S_m\}$ is the direct observation that the state of agent $n$ is in $\mathcal S_m$ at time $t$. During each rollout $k$, agent $n$ keeps the series of estimates $R_{\tau,t}^{n},\ \tau=kT_0,\ldots,(k+1)T_0-1 $, updating them recursively via \eqref{eqn_gossip_r} at each time $t=\tau,\ldots,(k+1)T_0-1$.
Using these \emph{gossiped} reward estimates to update the multipliers, the distributed version of \eqref{eqn_contractive_big_brother} becomes
\begin{align}\label{eqn_stochastic_dual}
    \hspace{-4pt}\lambda_{k+1}^n\hspace{-3pt}=\hspace{-3pt}\left[\hspace{-2pt}(1-\alpha)\lambda_k^n\hspace{-2pt}+\hspace{-2pt}\frac{\eta}{T_0}\hspace{-6pt}\sum_{\tau=k T_0}^{(k+1) T_0-1}\hspace{-14pt}\left(c-R_{\tau,(k+1)T_0-1}^n \right)\right]_{+}
\end{align}
where agent $n$ waits  until the end of the rollout at time $t=(k+1)T_0-1$ to substitute $R_{\tau,(k+1)T_0-1)}$ for $r(S_{\tau})$ and thus have the best possible estimate.  

Equations \eqref{eqn_gossip_r} and \eqref{eqn_stochastic_dual}, together with the policy structure in \eqref{eqn_separate_policy},  are the main tools for the distributed implementation of the dual update, allowing agents to run the fully distributed policy in Algorithm \ref{algo:alg_main_algorithm}.

In executing Algorithm \ref{algo:alg_main_algorithm}, agents use a \emph{realizable} policy $\pi_{\theta}[\lambda^1,\ldots,\lambda^N]$ in the sense that it is parametric, trained with finite rollouts as in \eqref{eqn_optimal_trained_policy}, and with local copies of the multipliers. Specifically, each agent $n$ substitutes its local copy of the vector of multipliers $\lambda^n$ for $\lambda$ in \eqref{eqn_separate_policy} to obtain  its local part of the realizable policy  $\pi_{\theta^n}(A_t^n\mid S_t^n,\lambda^n)$. Accordingly, the realizable global policy corresponds to  
\begin{align}\label{eqn_realizable_policy}
\pi_\theta[\lambda^1,\ldots,\lambda^N](A_t\mid S_t)&:=\prod_{n=1}^N\pi_{\theta^n}(A_t^n|S_t^n,\lambda^n).
\end{align}
By using the  distributed policies in \eqref{eqn_realizable_policy}, the team achieves the following realizable value
\begin{align}\label{eq:non-idealities-VR}
\hspace{-2pt}V_R&:=V_{T_0}\left(\pi_\theta[\lambda_k^1 \ldots \lambda^N]\right)\hspace{-2pt}
=\hspace{-2pt}\mathbb E_{\pi_\theta}\hspace{-2pt}\left[\frac{1}{T_0}\hspace{-4pt}\sum_{t=k T_0}^{(K+1) T_0-1}\hspace{-15pt}r(S_t)\hspace{-4pt}\ \right]\hspace{-2pt}.
\end{align}
Because agents cannot guarantee to reach exact consensus on the multipliers $\lambda^n$ almost surely when communicating over the stochastic graph \eqref{eqn_graph_model}, their realizable policies $\pi_{\theta^n}(\lambda^n)$ will differ from the one $\pi_{\theta^n}(\lambda)$ trained for, which assumed global access to $\lambda$. To mitigate the effect of this mismatch, we need the following assumption. 

\begin{assumption}($V$ functions Lipschitz continuity)\label{assumption:lipschitz}
The difference between the value functions $V_P$ and $V_R$ with global and distributed multipliers, as defined in \eqref{eq:non-idealities-VP} and \eqref{eq:non-idealities-VR}, respectively, is bounded by the scaled difference  their multipliers with scaling constant $L>0$, i.e., 
\begin{align}
\nonumber \|V_P-V_R\|&=\|V_{T_0}(\pi_\theta[\lambda, \ldots, \lambda]) - V_{T_0}(\pi_\theta[\lambda^1, \ldots, \lambda^N])\|\\& \leq L \max_{n=1,\ldots,N}\|\lambda - \lambda^n\|_\infty.
\end{align} 
\end{assumption}
This assumption requires the parametric policies in \eqref{eqn_optimal_trained_policy} to be Lipschitz with respect to the multipliers, which is the case, for instance, if we choose neural networks for the policy parameterization with bounded weights and Lipschitz activation functions such as ReLU.

\begin{algorithm}[t]
\caption{Distributed multi-agent  policy execution}\label{algo:alg_main_algorithm}
\begin{algorithmic}[1]
\FOR{$k=0,1,\ldots$}
  \STATE Substitute $\lambda^{k}_n$ in \eqref{eqn_optimal_trained_policy}  pretrained with local gradients.
  \STATE Keep the local realizable policy $\pi_{\theta^n}[\lambda^{k}_n]$ as in \eqref{eqn_realizable_policy}.
  \FOR{$t=k T_0,\ldots,(k+1)T_0-1$}
  	\STATE Act $A^n_t\sim \pi_{\theta^n}[\lambda^{k}_n]$ and transition to $S_{t+1}^n$.\\
    \STATE Collect local rewards $(R_{t,t}^n)_m=\mathds{1}\{S_t^n\in\mathcal S_m\}$. \\
  	\STATE \emph{Network gossiping:} Update $ R_{\tau,t}^n$ as in \eqref{eqn_gossip_r}.
  \ENDFOR
   \STATE Update the multipliers according to \eqref{eqn_stochastic_dual}.
\ENDFOR
\end{algorithmic}
\end{algorithm}

To quantify such an error, we measure the probability that the true rewards reach all agents. Referring back to Figures~\ref{fig:gossip_stochastic} and \ref{fig:timeline}, the event that at time $t$ agent $n$ knows whether there is an agent $n^\prime$ that satisfied $S_t^{n^\prime}\in \mathcal S_m$ at a previous time $\tau$  is given by $(R_{\tau,T}^n)_m = (r(S_{\tau}))_m$. The distribution of that event follows a negative binomial distribution, as is stated in the following proposition. 

\begin{proposition} \label{prop_binomial}
    Given the Bernoulli graph model \eqref{eqn_graph_model}, the probability that agent $n$ knows the global reward $r(S_{\tau})$  at time $t\geq \tau$ is
    $$\mathbb P\left((R_{\tau,t}^n)_m = (r(S_{\tau}))_m\right)\leq \mathbb P\left(BN(d_G, p)\leq t-\tau\right),$$
    where $d_G$ is the diameter of the underlying graph $G$, $p$ is the probability that an edge of the graph is active, and $BN(d_G,p)$ denotes the negative binomial distribution.
\end{proposition}
\begin{proof}
     See Appendix  \ref{app:proof_prop_binomial}.
\end{proof}
Leveraging Proposition \ref{prop_binomial}, the following result  provides a bound for the multipliers' error.
\begin{proposition}\label{prop_gossip_error}
       Given the Bernoulli graph model \eqref{eqn_graph_model}, the  error norm between multipliers $\lambda_k$ and $\lambda_k^n$ in \eqref{eqn_contractive_big_brother} and \eqref{eqn_stochastic_dual}, computed in terms of the rewards $r(S_\tau)$ and their estimators $R_{\tau,t}^n$, respectively is bounded by
\begin{align}\label{eq:lambdadif}\mathbb 
E\left[ \|\lambda_k^n-\lambda_k\|_\infty \right]\leq \frac{\eta d_G}{\alpha T_0 p},
\end{align}
where $d_G$ is the diameter of the underlying graph $G$, and $p$ is the probability that an edge of the graph is active.
\end{proposition}
 \begin{proof}
     See Appendix \ref{app:proof_prop_gossip_error}.
 \end{proof}
 
{According to \eqref{eq:lambdadif}, the error between the global and local multipliers is controlled by the step size $\eta$, the contraction parameter $\alpha$, and the rollout horizon  $T_0$.  
The trade-off for choosing $T_0$  entails keeping the truncation error $\epsilon$ in Assumption \ref{assumption_representation} sufficiently small and reducing the multipliers' error while ensuring that the multipliers are updated sufficiently often in order to speed training and increase the agents' agility in the online phase. Reducing $\eta$ also reduces the bound in \eqref{eq:lambdadif} at the expense of a slower dual update and policy reaction. 
}

\begin{remark}
    Proposition \ref{prop_gossip_error} also demonstrates the role of $\alpha$.  Including the contraction factor $(1-\alpha)$ in the dual update is key to guarantee that the deviation of the distributed multipliers remains bounded. Indeed, as the recursive addition of rewards accumulates over successive rollouts, the reward errors need to be modulated by powers of $(1-\alpha)$ to ensure that their aggregate effect in \eqref{eq:lambdadif} is subdued. A higher $\alpha$ makes the dual update more contractive, resulting in a smaller error in \eqref{eq:lambdadif}. However, increasing $\alpha$ has an adverse effect on the feasibility guarantees, as we will establish in the next section.
\end{remark}

\section{Feasibility Analysis}\label{sec:feasibility}

In this section we present the main result of our work which is to guarantee that the trajectories generated by Algorithm \ref{algo:alg_main_algorithm} are almost surely feasible in the sense that the time-averaged rewards exceed thresholds $c$, within an error controlled by the parameter $\alpha$. 

\subsection{Main Result}\label{ssec_main_result} 

Before stating the feasibility guarantees we require an additional assumption regarding the underlying MDP.

\begin{assumption}\label{assumption_noforces}(Controlled-invariant regions).
Every region $\mathcal S_m$ is controlled-invariant for each agent: for each $s\in \mathcal S_m$ there exists an action $a_m(s)\in\mathcal A$ that renders $\mathcal S_m$ invariant under the transition kernel, i.e., for all $t=0,1,2,\ldots$, $n=1,\ldots,N$, and $s\in \mathcal S_m$,
$$P(S^n_{t+1}\in \mathcal S_m \mid S^n_{t}=s, A^n_t=a_m(s))=1.$$
\end{assumption}
This invariance implies that, if necessary, an agent can keep its state stationed in a particular region $S_{t}^n\in \mathcal S_m$. 
\textcolor{black}{We add this assumption to ensure feasibility with a margin and simplify the statement and proof of Lemma \ref{lemma:positive_ideal_value} on the next page. However, the strict invariance is not fundamental. It could be substituted by a milder version in which an agent is capable of keeping its state in $\mathcal S_m$ for only a fraction of time long enough such that the problem remains feasible. This would only affect the parameter $\delta_\alpha$ our main result, which we state next.}


\begin{theorem}\label{theorem:feasibility}
Let Assumptions \ref{assumption_representation}--\ref{assumption_noforces} hold, and consider the specifications $\|c\|_\infty <1$,  and $\|c\|_1 \leq N-1$. If 
\begin{align}\label{eq:delta_theorem}
\delta:=1-\|c\|_\infty - M\left(\epsilon+\beta\right)>0,
\end{align} 
the trajectories generated by Algorithm \ref{algo:alg_main_algorithm} over the stochastic  communication network \eqref{eqn_graph_model}  are feasible within an error  $\sqrt{\alpha M}$ with probability one, i.e.

\begin{equation}\label{eqn_as_feasibility}
    \liminf_{T\to \infty}\frac{1}{T} \sum_{t=0}^{T-1}  r(S_t)\geq c - \mathbf  1 \sqrt{\alpha M}, \mbox{ a.s.}
    \end{equation} 
with $\mathbf 1$ being the vector of all ones and 
\begin{equation}\label{eq:alpha_bound}
    \alpha\geq\frac{\eta d_G}{p T_0}  \frac{M L}{\delta}.
\end{equation}
\end{theorem}
\begin{proof}
    See Section \ref{ssec:proof_th}.
\end{proof}

Since the constraints  \eqref{eqn_crl} involve time averages of binary rewards, the thresholds in $c$ must be no larger than one for a feasible policy to exist. Theorem \ref{theorem:feasibility} imposes a stricter requirement $\|c\|_\infty <1$, 
leaving a margin for the realizable policies.  Also  $\|c\|_1 \leq N-1$ is imposed, ensuring that $N-1$ agents can satisfy all constraints with a margin for the transients when agents' states cycle around regions $\mathcal S_m$. Additionally, condition \eqref{eq:delta_theorem} 
imposes limits on the approximation error in training, with $\beta$ and $\epsilon$ being the quantities defined in Assumption \ref{assumption_representation}.
With $\|c\|_\infty$ representing the stringest requirement, which can be satisfied according to   Lemma \ref{lemma:positive_ideal_value} by a full policy \eqref{eqn_ideal_separated_policy} producing values $V$, the extra slack  $M(\beta+\epsilon)$ is required because these values are degraded to $V_R$ when running the realizable policy \eqref{eqn_realizable_policy} instead. 
Regarding the requirement for $\alpha$ in \eqref{eq:alpha_bound}, notice that it can be made arbitrarily small by designing a rich parameterization or neural network, a long time horizon $T_0$, and a small step size $\eta$. These parameters introduce a trade-off between reducing the error and slowing training. 

The result in Theorem \ref{theorem:feasibility} requires a slack margin of $\sqrt{\alpha M}$ on the requirements specified by $c$. 
Alternatively, strict feasibility for problem \eqref{eqn_crl} is achieved if we run our algorithms with the augmented parameter $ c_\alpha = c + \sqrt {\alpha M}$.  We formalize this claim in the following corollary.
\begin{corollary}
Consider problem $\eqref{eqn_crl}$  with $\|c\|_\infty  \leq 1- \sqrt{\alpha M}$, and  $\|c\|_1+M^{3/2}\alpha \leq N-1$   where  $\alpha$ is a constant satisfying  $\alpha\geq\frac{\eta d_G}{p T_0}  \frac{M L}{ \delta_\alpha},$ and $ \delta_\alpha:=1-\|c\|_\infty-\sqrt{\alpha M}- M\left(\epsilon+\beta\right)>0$. \\
Under Assumptions \ref{assumption_representation}--\ref{assumption_noforces},  the trajectories  
the trajectories generated by Algorithm \ref{algo:alg_main_algorithm} over the stochastic  communication network \eqref{eqn_graph_model}  with multipliers updated by
\begin{align*}\label{eqn_stochastic_dual}\lambda_{k+1}^n\hspace{-2pt}=\hspace{-2pt}\left[\hspace{-2pt}(1-\alpha) \lambda_k^n + \frac{\eta}{T_0}\hspace{-5pt}\sum_{\tau=k T_0}^{(k+1) T_0-1}\hspace{-5pt}\left(\hspace{-1pt}c\hspace{-2pt}+\hspace{-2pt}\sqrt{\alpha M}\hspace{-2pt}-\hspace{-2pt}R_{\tau,(k+1)T_0-1}^n\hspace{-2pt} \right)\hspace{-2pt}\right]_{+}
\end{align*}
are feasible  with probability one, i.e.
\begin{equation}\label{eqn_as_feasibility_corollary}
    \liminf_{T\to \infty}\frac{1}{T} \sum_{t=0}^{T-1}  r(S_t)\geq c, \mbox{ a.s.}.
    \end{equation} 
\end{corollary}
Similar to those in  Theorem \eqref{theorem:feasibility}, the conditions on $\alpha$ and $\delta_\alpha$ in this corollary can be met, provided that $\|c\|_\infty<1$, by designing the stepsize, horizon, and parameterization so that  $\eta/T_0$ and $\epsilon+\beta$ become sufficiently small. 

The result in Theorem \ref{theorem:feasibility} is derived by unrolling the stochastic dual recursion in \eqref{eqn_stochastic_dual} to write the sum of the rewards in terms of the multipliers, which yields an expression for the feasibility error in terms of the stochastic average of the multipliers along a trajectory. 
Hence, the proof is completed by bounding this stochastic average by $\eta \sqrt{M/\alpha}$. 
To obtain this bound for the stochastic average, we utilize proof techniques similar to those in stochastic gradient descent on the dual domain,
 for which it is key to show that the gradient of the dual forms an acute angle with the vector of multipliers, i.e.    
\begin{align}
\lambda^\top (V-c)&\geq 0.\label{eqn_accute_angle}
\end{align}
In this direction, Lemma \ref{lemma:positive_ideal_value} below indicates that if we use the ideal policy $\pi^\star[\lambda,\ldots,\lambda]$ in \eqref{eqn_optimal_set}, the condition  \eqref{eqn_accute_angle} is satisfied even with a slack  $(1-\|c\|_{\infty})\|\lambda\|/\sqrt{M}$, that will become handy when considering the realizable policy in \eqref{eq:non-idealities-VR} instead.  The proof of Lemma \ref{lemma:positive_ideal_value} uses Assumption \ref{assumption_noforces} to ensure that the underlying MDP does not prevent the agents from taking care of the most pressing constraints. The value $V$ in \eqref{eqn_ideal_V} and \eqref{eqn_accute_angle} assumes that we could train with infinite horizon, without the limitations of parametrization, and with all agents having access to the same  multipliers in \eqref{eqn_contractive_big_brother}. On the other end, $V_R$ in \eqref{eq:non-idealities-VR} corresponds to the  realizable policy in Algorithm \ref{algo:alg_main_algorithm},  and is obtained by  applying  the distributed, finite horizon  and parametric procedures behind equations \eqref{eqn_optimal_trained_policy} and \eqref{eqn_stochastic_dual}. In particular, each agent utilizes its own copy of local multipliers in \eqref{eqn_stochastic_dual}. In between $V$ and $V_R$, we defined  $V_P$ in \eqref{eq:non-idealities-VP} as a theoretical value to be used as an intermediate step in the proofs of feasibility. It differs from $V$ in using a parametric policy and in aggregating the rewards for a finite time period of length $T_0$. And $V_P$ is different from $V_R$ in assuming global access by all agents to the same multipliers. We use Assumptions \ref{assumption_representation} and \ref{assumption:lipschitz} in Lemma \ref{lemma:error_gradient} to bound the difference $\lambda^\top(V-V_R)$. If this difference is lower than the slack $(1-\|c\|_{\infty})\|\lambda\|/\sqrt{M}$, then \eqref{eqn_accute_angle} will still be satisfied if we substitute  $V_R$ for $V$. 

\subsection{Proof of Theorem \ref{theorem:feasibility} and Supporting Results}\label{ssec:proof_th}

We proceed to prove our first result. The intuition in Lemma \ref{lemma:positive_ideal_value} is that if we fix $\lambda$, the ideal optimal policy assigns the $N$ agents to the regions $\mathcal S_m$ with highest weighted rewards. As the multipliers cycle in the execution phase according to Algorithm \ref{algo:alg_main_algorithm}, this ideal behavior will drive the agents to cyclically attend the most pressing constraints at each rollout.   

\begin{lemma}\label{lemma:positive_ideal_value}
Assumption \ref{assumption_noforces} with $\|c\|_{\infty}<1$ and $\|c\|_1 \leq N-1$  as in Theorem \ref{theorem:feasibility} results in
 \begin{align}\lambda^\top (V-c)&\geq (1-\|c\|_{\infty})\frac{\|\lambda\|}{\sqrt{M}}.\label{eqn_lemma_vI}
 \end{align}
 \end{lemma}
\begin{proof}
Without loss of generality, sort the multipliers $\left(\lambda\right)_1\geq \left(\lambda\right)_2\geq\cdots\geq \left(\lambda\right)_M$, so that $\left(\lambda\right)_1$ is the largest multiplier.

One policy that optimizes the Lagrangian would assign an agent to each region $\mathcal S_m$ of the state space for $m=1\ldots,N$, since there are only $N<M$ agents. Therefore, $\left(V\right)_m=1$ for $m\leq N$ and $\left(V\right)_m=0$ for $m>N$. Hence it follows that
\begin{align}\label{eq:inner_product_decomposition}
\lambda^\top (V-c) &= \left(\lambda\right)_1 (1-\left(c\right)_1) \\
&+ \sum_{m=2}^N \left(\lambda\right)_m (1-\left(c\right)_m) 
+\hspace{-9pt}\sum_{m=N+1}^M\hspace{-5pt}\left(\lambda\right)_m (0-\left(c\right)_m). \notag
\end{align}
Next, we use the fact that $\left(c\right)_1 \leq \|c\|_{\infty}$ by definition of the infinity norm, so that $(1-(c)_1)\leq (1-\|c\|_\infty)$. Also, the multipliers were sorted in decreasing order which implies that the first one is maximum $(\lambda)_1=\| \lambda\|_\infty$, the following  $N-1$  satisfy $ (\lambda)_m\geq (\lambda)_N$ for all $2\leq m\leq N$, and the last $M-N$ satisfy $(\lambda)_m\leq (\lambda)_N$ for all $m>N$. Hence, we bound the three terms in  \eqref{eq:inner_product_decomposition} by
\begin{align}
\lambda^\top (V-c) &\geq (1-\|c\|_{\infty}) \|\lambda\|_\infty\\ &+(\lambda)_N  \sum_{m=2}^N (1-\left(c\right)_m)
-(\lambda)_N  \sum_{m=N+1}^M \left(c\right)_m\nonumber\\
&\geq (1-\|c\|_{\infty}) \|\lambda\|_{\infty} +(\lambda)_N \left((N-1) - \|c\|_1\right), \nonumber
\end{align}
where we grouped the terms multiplying $\lambda_N$ and then used $\|c\|_1=\sum_{m=1}^M (c)_m\geq \sum_{m=2}^M (c)_m$. Given that  $(N-1) \geq \|c\|_1$ by hypothesis, we can write
\begin{align}
\lambda^\top (V-c) &\geq (1-\|c\|_{\infty}) \|\lambda\|_{\infty}. 
\end{align}
Then use the property $\|\lambda\|_{\infty} \geq \|\lambda\|/\sqrt{M}$.
\end{proof}

The following results bounds the error incurred in executing the realizable policy \eqref{eqn_realizable_policy} as compared to the ideal policy \eqref{eqn_ideal_separated_policy}. Intuitively, Lemma \ref{lemma:error_gradient} guarantees that when agents use the realizable policy, they still behave as in the ideal case of  Lemma \ref{lemma:positive_ideal_value}, attending the most pressing constraints during the current rollout. Its proof uses Assumption  \ref{assumption_representation} to bound the error between $V$ and $V_P$, and Assumption \ref{assumption:lipschitz} together with Proposition \ref{prop_gossip_error} for the error between $V_P$ and $V_R$. 

\begin{lemma}\label{lemma:error_gradient}
Given the stochastic communication network model \eqref{eqn_graph_model}, 
Assumptions \ref{assumption_representation}--\ref{assumption:lipschitz} ensure that the following bound holds for the difference between the gradient of the ideal and realizable value functions %
\begin{align}  (\lambda_k)^\top \left(V-V_R\right)\leq \|\lambda_k\|\left(\epsilon+\beta+\eta L \sqrt{M} \frac{d_G}{\alpha T_0 p}\right),\label{eqn_lemma_Error_VI_VR}
\end{align}
where $V$ and $V_R$ are defined in \eqref{eqn_ideal_V} and \eqref{eq:non-idealities-VR}.
\end{lemma}
\begin{proof}

 Obtain the bound $
    \left\|V_R-V\right\| \leq\left\|V-V_P\right\|+\left\|V_P-V_R\right\|
$
by adding and subtracting the term $V_P$ and using the triangle inequality.  Hence, by  Assumptions \ref{assumption_representation} and \ref{assumption:lipschitz} 
$\left\|V_R-V\right\| \leq \epsilon+\beta+L \max_{n=1,\ldots,N} \left\|\lambda_k^n-\lambda_k\right\|_\infty,
$
and lastly, using Proposition \ref{prop_gossip_error}, we obtain the result
$\left\|V_R-V\right\| \leq \epsilon+\beta+L \epsilon_G=\epsilon+\beta+\eta L \sqrt{M} d_G/(\alpha T_0 p).$
\end{proof}
Combining Lemmas \ref{lemma:positive_ideal_value} and  \ref{lemma:error_gradient}, if conditions \eqref{eq:delta_theorem} and \eqref{eq:alpha_bound} in Theorem \ref{theorem:feasibility} are satisfied, then \eqref{eqn_accute_angle} holds for the realizable policy and $V_R$. Indeed, from \eqref{eqn_lemma_vI} and \eqref{eqn_lemma_Error_VI_VR}  it yields 
\begin{align}
&\lambda^\top (V_R-c) = \lambda^\top(V-c) + \lambda^\top (V_R-V)\label{eqn_real_accute}\\
&\geq\frac{\|\lambda\|}{\sqrt{M}}\left((1-\|c\|_\infty)-M\left(\beta+\epsilon +\eta L  \frac{d_G}{\alpha T_0 p}\right)\right)\geq 0 ,  \nonumber 
\end{align}
where the last inequality requires $\delta>0$ in \eqref{eq:delta_theorem} and it is equivalent to $\alpha \geq {\eta d_G M L}/(p T_0 \delta )$ in \eqref{eq:alpha_bound}.

Having a positive inner product in \eqref{eqn_real_accute}, we can prove that the stochastic average of the multipliers across a trajectory is bounded.  
This key result is presented below as Proposition \ref{proposition:combined_lemma}.  Its proof,  relies on three steps. First, we combine \eqref{eqn_real_accute} with the update \eqref{eqn_contractive_big_brother} to obtain a contractive system with constant input for the expected value of $\|\lambda\|$ conditioned on past values. From this contraction, it follows that the expected value of the time average of the multipliers is bounded. Lastly, we include a modified proof of the strong law of large numbers for non-i.i.d. variables, which lets us drop the expected value and reach the following result.

\begin{proposition}
\label{proposition:combined_lemma}
Under Assumptions \ref{assumption_representation}--\ref{assumption_noforces}, given the  stochastic  communication network model \eqref{eqn_graph_model},  and with $\|c\|_\infty<1$, $ \|c\|_1\leq N-1$, $\alpha\geq\frac{\eta d_G}{p T_0} \cdot \frac{\sqrt{M} L}{\delta}$, and 
$\delta=\left(1-\|c\|_\infty\right) - \sqrt{M}\left(\epsilon+\beta\right)>0,$  the averaged rewards along a trajectory are bounded by $\eta \sqrt{M/\alpha}$ with probability $1$, i.e.,
\begin{align}\label{eqn_bound_stochasstic_average}
\limsup_{k \rightarrow \infty} \frac{1}{k} \sum_{i=1}^{k} \left(\lambda_i\right)_m \leq \eta \sqrt{\frac{ M}{\alpha}} \text { a.s.. }
\end{align}
\end{proposition}
\begin{proof}
See Appendix \ref{app:combined_lemma}.
\end{proof}

\begin{figure*}[t!]
   \centering
   \begin{subfigure}{0.65\columnwidth}
   		\centering
        \includegraphics[scale=1]{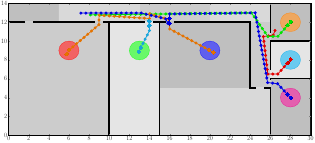}        
   		\caption{Floor plan.}	
            \label{fig:floorplan}
   \end{subfigure}
   \begin{subfigure}{0.65\columnwidth}
   		\centering
        \includegraphics[scale=1]{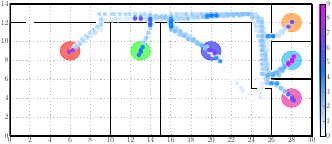}    
   		\caption{Agent $0$.}
   		\label{fig:heatmap_agent0}
   \end{subfigure}
   \begin{subfigure}{0.65\columnwidth}
   		\centering
        \includegraphics[scale=1]{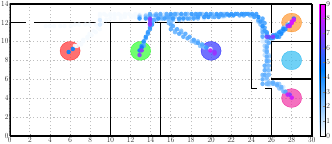}            
   		\caption{Agent $1$.}
   		\label{fig:heatmap_agent1}
   \end{subfigure}   \\
   \begin{subfigure}{0.65\columnwidth}
   		\centering
        \includegraphics[scale=1]{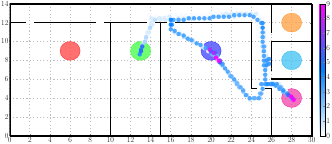}    
        \caption{Agent $2$.}
   		\label{fig:heatmap_agent2}
   \end{subfigure}   
     \begin{subfigure}{0.65\columnwidth}
   		\centering
        \includegraphics[scale=1]{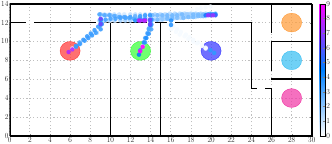}    
   		\caption{Agent $3$.}
   		\label{fig:heatmap_agent3}
   \end{subfigure}
   \begin{subfigure}{0.65\columnwidth}
   		\centering
        \includegraphics[scale=1]{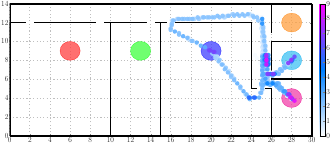}    
   		\caption{Agent $4$.}
   		\label{fig:heatmap_agent4}
   \end{subfigure}   
\caption{(a) Floor plan and sample trajectories for each of the $N=5$ agents, $n=0,\ldots,4$. Black lines represent walls, and colored circles represent the $M=6$ zones to be patrolled by the agents. Each colored dot along a trajectory indicates a new position for each time step. The $12$ gray rectangular regions define distinct possible tiles or observations of an agent’s position, which are the inputs to the policy. Figures (b)--(f): Occupation heat maps. The complete trajectory of an agent across $40{,}000$ timesteps is represented by dots, indicating the position reached by an agent. Each dot color represents the frequency of occupation at that position, with darker hues indicating more frequent positions, under a logarithmic colorbar scale.}
\label{fig:heatmaps}
\end{figure*}
Hence, we proceed with the proof of  Theorem \ref{theorem:feasibility}. It yields from applying the results of Proposition \ref{proposition:combined_lemma} to the update rule \eqref{eqn_contractive_big_brother}.  Specifically, we introduce
\begin{equation}
    \bar r_k^{T_0}=\frac{1}{T_0}\sum_{\tau=(k-1)T_0}^{kT_0-1}r(S_t),
\end{equation}
which  reduces notation and lets us rewrite \eqref{eqn_contractive_big_brother} as
\begin{align}\label{eq:proyection_inequality}
    \lambda_{k+1}&=\left[(1-\alpha) \lambda_k+\eta\left(c-\bar r_k^{T_0}\right)\right]_{+}\\
    &\geq (1-\alpha)\lambda_k+\eta\left(c-\bar r_k^{T_0}\right), \notag
\end{align}
where the inequality must be understood component-wise. Thus, from \eqref{eq:proyection_inequality}, we can rewrite
\begin{align}&
\eta\left(\bar r_k^{T_0}-c\right) \geq  (1-\alpha) \lambda_k -\lambda_{k+1}.
\end{align}
Hence, assuming $\lambda_0=0$, adding over $k=0,\ldots,K$
\begin{align}\label{eq:constraint_lambda}
&\sum_{k=0}^K \eta\left(\bar r_k^{T_0}-c\right) \geq \sum_{k=0}^K(1-\alpha) \lambda_k-\sum_{k=0}^K\lambda_{k+1} \\
&
= \sum_{k=1}^K(1-\alpha)  \lambda_k  -\sum_{k=1}^{K} \lambda_{k} -\lambda_{K+1} 
= -\sum_{k=1}^K\alpha  \lambda_k   -\lambda_{K+1}, \nonumber 
\end{align}
where we removed the term corresponding to $\lambda_0$, substituted $k=k+1$ in the sum of $\lambda_{k+1}$ and then rearranged two sums into one removing those terms $\lambda_k$ that cancel each other.  We further obtain  
\begin{align}
    & \frac{1}{K} \sum_{k=0}^K  \bar r_k^{T_0}\geq c -  \frac{1}{\eta K} \lambda_{K+1}  - \frac{\alpha}{\eta}\frac{1}{K} \sum_{k=1}^{K}   \lambda_k, \label{eq:constraint_with_lambdaK}
\end{align}
 by moving $c$ and $\eta$ to the right-hand side, and then use Proposition \ref{proposition:combined_lemma} to arrive at
\begin{align}& \liminf_{K\to\infty}\frac{1}{K} \sum_{k=1}^K  \bar r_k^{T_0}\geq c-\alpha\sqrt{\frac{ M}{\alpha}}.  \label{eq:constraint_satisfaction_mean}
\end{align}
We demonstrated throughout this section that selecting $\alpha>0$, which is required according to Proposition \ref{prop_gossip_error} to ensure that the multiplier error remains bounded, introduces an error in the feasibility result \eqref{eqn_as_feasibility}. Thus, we want to make $\alpha$ as small as possible, but this is limited by \eqref{eq:alpha_bound}, which is imposed to ensure the dual descends according to  \eqref{eqn_real_accute}. We can adjust $\eta$ and $T_0$ and $p$ to reduce the lower bound for $\alpha$ in \eqref{eq:alpha_bound}. In particular, the activation probability $p$ in the stochastic communication graph depends on the communication scheme that is part of the problem design. In the next section, we provide an example of how to increase this probability by boosting the communication power.  

\section{Numerical Experiments}\label{sec_numerical}
In the following, we present numerical experiments designed to test the performance of the proposed multi-agent reinforcement learning Algorithm \ref{algo:alg_main_algorithm}. We consider a scenario with five robots acting as agents, which navigate and monitor  multiple rooms in a floor plan.
    
\subsection{Floor Plan Navigation}

Consider $N=5$ agents navigating a floor plan of an L-shaped corridor connecting three offices and three laboratories at the University of Pittsburgh, with their separating walls represented as black lines (see Fig. \ref{fig:heatmaps}). In this scenario, the state of agent $n$ at time $t$ is $S^n_t=(x^n_t,y^n_t)$  with $x^n_t$ and $y^n_t$ representing the agent horizontal and vertical coordinates, respectively with $S^n_t \in [0,30]\times[0,14]$, all measured in meters. Correspondingly, the $M=6$ regions    $\mathcal S_1,\ldots,\mathcal S_M$ are depicted by the colored circles in Fig. \ref{fig:heatmaps}, centered at $(x_1,y_1)=(6,9)$, $(x_2,y_2)=(13,9)$, $(x_2,y_2)=(20,9)$, $(x_3,y_3)=(28,4)$, $(x_4,y_4)=(28,8)$, and $(x_5,y_5)=(28,12)$, and with all  radius equal to one. The global reward corresponding to zone $m$ takes the value $(r(S_t))_m=1$ if at least one agent enters the corresponding circle. The action space $\mathcal A=\{0,\ldots,5\}$ is finite, meaning that the agents must decide  which of the six regions to visit at each instant. We assume a low-level robot  navigation control is available to drive the agent across the floor plan toward the selected zone. This controller follows a sequence of segments between intermediate goals located at a set of strategic points in rooms, corridors, and doors connecting any two rooms in the floor plan of Fig. \ref{fig:heatmaps}. 

Next, we describe in more detail how agents gain this ability to identify which zones they should visit by applying Algorithms \ref{alg:offline} and \ref{algo:alg_main_algorithm}. First, each agent learns a set of optimal policy parameters offline using Algorithm \ref{alg:offline}. Specifically, each agent adopts a soft-max policy with exponents given by the $M=6$ logits in the output layer of a two-layer neural network with $256$ hidden neurons whose inputs are the vector $\lambda$ of $M=6$ multipliers and the coordinate $S_t^n$ of the agent.

To optimize \eqref{eqn_optimal_trained_policy} for all agents efficiently, we take a block maximization approach in which individual agents retrain sequentially. We start this procedure by training a single agent in the floor plan environment to obtain a set of parameters $\bar \theta$. Then, all $N$ agents' policies are initialized with the same common parameter $\theta^n=\bar\theta,\ n=1,\ldots,N$, and the retraining sequence starts. In each step of this sequence, an agent is selected to be the active learner, which updates its policy by running the iteration in Algorithm \ref{alg:offline}. All other agents follow trajectories driven by their policies, which remain unchanged. After the active agent retrains in these conditions, its policy settles, and the active token is passed on to the next agent. 
   
During the online execution phase,  all agents initialize their multipliers to zero and then update them following \eqref{eqn_stochastic_dual} with $\alpha=0.01$ and $\eta=0.5$. For a span of $T_0=100$ time steps, each agent uses the realizable policy described in   Algorithm \ref{algo:alg_main_algorithm}. They substitute their local copies $\lambda_k^n\in\mathbb R^M$  in the trained policy $\pi_\theta(S_t,\lambda_k^n)=(\pi_{\theta^1}(S_t^1,\lambda_k^n),\ldots,\pi_{\theta^N}(S_t^N,\lambda_k^n))$, and then keep the $n$-th component   $\pi_{\theta^n}(S_t^n,\lambda_k^n))$. Using this procedure, the agents do not need to know or exchange their local positions. Instead, the coordination is achieved through the dual multipliers \eqref{eqn_stochastic_dual}, which are kept consistent among agents thanks to the network gossiping of reward estimations  \eqref{eqn_gossip_r}. As detailed in Section\ref{ssec:numerical_stochastic_graph} below, the agents in this experiment gossip through an ad-hoc communication network that resembles the probabilistic model \eqref{eqn_graph_model}, in which two agents communicate if they are inline of sight or near each other.   

The coordinated behavior in which the agents take complementary paths is observed in Fig. \ref{fig:heatmaps}, where we show $N=5$ heat maps, each corresponding to the trajectories of an agent during the online execution phase.   Specifically, each of the $N=5$ color graphs in  Fig. \ref{fig:heatmaps}, one per agent,  is obtained by dividing the floor plan into $120\times56$ square bins of side $dx=dy=0.25$ and computing a two-dimensional histogram by counting how many times the agent enters each bin during a trajectory of $T=40,000$ time steps. The color scale is logarithmic,  which implies that the agents occupy most of their time inside the circular regions. The paths in Fig. \ref{fig:heatmaps} demonstrate how the agents learn to coordinate in order to attend different regions and thus satisfy the competing constraints. The constraints for this numerical experiment are defined by the thresholds $c=(0.1,0.2,0.3,0.4,0.5,0.6)$ for the red, green, blue, orange, cyan, and magenta regions, respectively. Since the sum of these thresholds is $\|c\|_1=2.1$ greater than one, a single agent working alone cannot satisfy the constraints. And because there are more zones $M=6$ than agents $N=5$ the constraints cannot be satisfied by placing an agent in each zone. By applying Algorithms \ref{alg:offline} and \ref{algo:alg_main_algorithm}, the agents learn to automatically achieve a high-level coordination in which agents $n=0,1,2,3$ and $4$ are sent to attend most of their time at the cyan, orange, blue, green and magenta zones, respectively, and they use their spare time to collaborate in satisfying the constraint corresponding to the red zone. The assignment is not trivial because it depends on the relative weights of the thresholds and the proximity of the zones. For instance, agent $n=3$ contributes most of its time to attending the red zone, but it is otherwise attending the green zone in the next room, so it is the closest one to help. Moreover, it is intuitive that the red and green zones can split an agent because they are associated with less demanding constraints.    

\begin{figure}[t]
    \centering
    \includegraphics[scale=1]{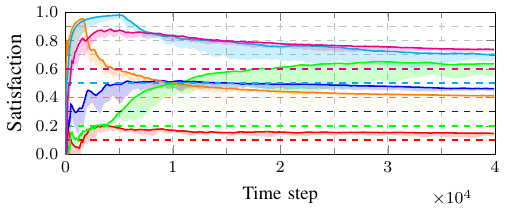}    
    \caption{Satisfaction of the constraints for each zone $m=1,\ldots,6$. Constraint requirements were defined as
$0.1$ for zone 1, $0.2$ for zone 2, etc. Dashed lines indicate these constraints. Minimum and maximum satisfaction values are plotted for each timestep for each zone and are filled by a shaded region. Colors of constraint and satisfaction values match those of the zones as depicted in Fig. \ref{fig:heatmaps}.}
    \label{fig:satisfaction_floorplan}
\end{figure}

\begin{figure*}[t!]
   \centering
   \begin{subfigure}{0.45\columnwidth}
  		 \centering
        \includegraphics[scale=1]{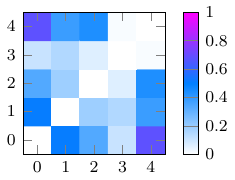}    
   		\caption{Communication matrix.}
   		\label{fig:gossip_matrix}
   \end{subfigure} 
   \begin{subfigure}{0.775\columnwidth}
   		\centering
        \includegraphics[scale=1]{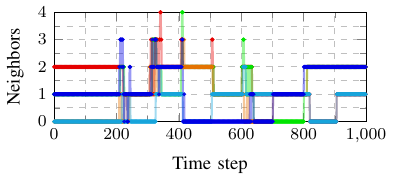}    
   		\caption{Communication neighborhoods.}
   		\label{fig:gossip_trajectories}
   \end{subfigure}
   \begin{subfigure}{0.775\columnwidth}
   		\centering
        \includegraphics[scale=1]{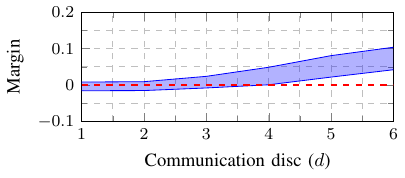}    
   		\caption{Constraint margins.}
   		\label{fig:communication_discs}
   \end{subfigure}
   \caption{
     (a) Matrix of communication frequencies between agents, counted over time. For each pair of
agents ($n$, $n^\prime$), the number of timesteps during which these agents communicate is summed and divided by the total timesteps of simulation, $40{,}000$. Frequencies are indicated by colors matching those in the included colorbar, with white indicating no communication. We define an agent as never communicating with itself, so diagonal entries are white. Figure (b): A snapshot of gossip neighborhood sizes per agent. Neighborhood sizes are plotted for $1{,}000$ time-steps of execution phase. $N=5$ lines are present, one for each agent and of a color matching said agent’s trajectory in Fig \ref{fig:heatmaps}. Figure (c): Margin of constraint satisfaction for each communication disc of sizes $d=1,\ldots,6$. A minimum and maximum difference between satisfaction and constraint values across all zones $m = 1,\ldots,6$ are found and plotted for each disc size, as indicated by the bold blue lines. A band shades the area between the maximum and minimum differences.}
\end{figure*}
   
Overall, the underlying mechanism agents use to coordinate is to let themselves be driven by the highest multipliers' values while avoiding flocking. The high-level control that drives the agents is akin to a standard dual-optimal dynamic, in which each multiplier (increases) decreases when its corresponding zone is (not) being attended, making it (more) less urgent for the agents to visit it. Indeed, the policies are trained to seek the highest weighted reward, $r_\lambda(S_t)=\lambda^\top(r(S_t)-c)=\sum_{m=0}^M(\lambda)_m((r(S_t)-c)_m)$ and this is achieved by setting to one the rewards $(r(S_t))_m$ corresponding to the highest multipliers, as it was argued for the ideal policy $\pi_I$ in the proof of Lemma \ref{lemma:positive_ideal_value}. 

Figure \ref{fig:satisfaction_floorplan} presents the resulting performance in terms of constraint satisfaction. The dashed horizontal lines in Fig. \ref{fig:satisfaction_floorplan} represent the thresholds $c$ the agents must comply with. The colors correspond to the zones in Figure \ref{fig:heatmaps}. The curves correspond to the long-run time averages $\bar r(S_t)=\frac{1}{t}\sum_{\tau=0}^{t-1}r(S_{\tau})$ over the  span of $40{,}000$ time steps. Notice that we use the global rewards $r(S_t)$ as compared to $R_{\tau,t}^n$, even though they are never computed by the agents when running Algorithms \ref{alg:offline} and \ref{algo:alg_main_algorithm}, because these are the ones defining our original optimization problem \eqref{eqn_crl}. Accounting for the randomness of the reward trajectories, we  run Algorithm \ref{algo:alg_main_algorithm} five times and report a band between the maximum and minimum values of the averaged rewards $\bar r(S_t)$.    The results demonstrate that the agents can meet the specified constraints consistently for all runs so that all zones are adequately monitored over time, outperforming numerically the almost sure theoretical guarantees provided by Theorem \ref{theorem:feasibility}. 

\subsection{Communication over a Stochastic Graph}\label{ssec:numerical_stochastic_graph}

Purposely, the communication setup in this experiment does not exactly match the stochastic graph model with Bernoulli edges in \eqref{eqn_graph_model}. Instead, we consider a practical ad-hoc network that changes its connectivity as agents move across the floor plan. Specifically, two agents communicate if they are in the line of sight of each other regardless of their distance, or if they are closer than a range disc size $d= 5$ meters separated by a wall. The fraction of time that each pair of agents communicate to each other along a trajectory of $40{,}000$ steps is depicted in Fig. \ref{fig:gossip_matrix}. This representation shows that agents $n=0$ and $n=4$ remain connected most of the time,  reflecting that they are within communication range, because they share similar paths and spend most of their time stationed in rooms next to each other. All agents connect to at least one other agent over time, enabling the gossip protocol \eqref{eqn_gossip_r} to succeed in passing the reward estimates between any two agents via multi-hop communications. Hence, the underlying static graph is connected, but not its stochastic samples, since the agent $n=3$ is frequently isolated from all other agents, probably when visiting the remote red zone, so that we would see a clustered network most of the time if we sampled the edges per time step. This intermittent communication connectivity plays a central role in the agents' behavior. For instance, Fig. \ref{fig:floorplan} shows the blue trajectory of an agent entering the third room from the left but changing its path towards the magenta zone once it forms a line of sight with the orange agent attending the blue zone.  

Another perspective of the network connectivity is given in Fig. \ref{fig:gossip_trajectories}, which shows how the sizes of the communication neighborhoods evolve across time during a span of one thousand timesteps. There are five lines in Fig. \ref{fig:gossip_trajectories}, representing the number of neighbors for each agent, with the same color-to-agent correspondence as in Fig. \ref{fig:floorplan}. These trajectories corroborate that the agents do not flock together since they isolate reaching zero neighbors part of the time and are rarely connected to two other agents. It also shows that they seldom confer altogether, only two times in a thousand, so they need to transmit the rewards across the stochastic network by moving and passing messages dynamically. We can also infer from Fig. \ref{fig:gossip_trajectories} that the agents cannot remain static in a zone during the total span of the experiment. The communication neighborhoods change dynamically as agents adjust their trajectories and spend part of their time for communication purposes, alternating between zones aiming to increase their joint reward and exchange information with their pairs.  This behavior aligns with the observation that none of the agents hold their position throughout the experiment in Fig. \ref{fig:heatmaps}, which is reasonable from the algorithmic perspective since an isolated agent sitting in a zone would see all other $M-1$  multipliers  growing, and it would be therefore inclined to move to attend their corresponding zones, leaving its own.    

Finally,  we analyze how varying the communication range affects performance. Fig. \ref{fig:communication_discs} examines the margin of constraint satisfaction relative to the size of the communication disc $d$. For $d=1,\ldots,6$,  the blue lines represent   the minimum difference between the average rewards and constraint thresholds across zones at time $t=40{,}000$. That is, $\min_m (\bar r_{40,000}-c)_m$. The experiment is repeated five times to account for the randomness in the trajectories, and a band between the maximum and the minimum margin is presented in Fig. \ref{fig:communication_discs}. 
   
The results show that the margin of constraint satisfaction results in negative values for small values of $d$, therefore not complying with primal problem \eqref{eqn_crl},  increasing above $0$ for when the communication range increases above $d=4$. This dependence of the margin with $d$ aligns with our theory in Theorem \ref{theorem:feasibility}, considering that a larger disc corresponds with a higher probability $p$. Thus, \eqref{eqn_as_feasibility} tells us that the error of feasibility becomes smaller when $\alpha$ decreases, which according to \eqref{eq:alpha_bound} is driven by increasing the probability $p$. Alternatively, for a fixed value of the parameter $\alpha$, equation \eqref{eq:alpha_bound} can only be satisfied if $p$ becomes big enough.
 
In summary, these numerical experiments illustrated how our coordinated offline training, dual dynamics, and gossiping protocol provide a system of multiple agents with the ability to coordinate their actions.      %
The simplified stochastic graph model for communication in \eqref{eqn_graph_model} has the virtue of being tractable, which allowed us to advance our analysis of feasibility summarized in Theorem \ref{theorem:feasibility} while  capturing the necessary properties of a more realistic distance-defined network in terms of probabilities and time-evolving connectivity. Hence, the experiments abide by the theory, so the numerical results corroborate  our theoretical feasibility guarantees.

\section{Conclusion}
\label{sec:conclusion}
In this work, we demonstrated the effectiveness of CMARL in complex coordination problems in which agents must solve competing tasks. By employing a state augmentation framework with non-convergent dual variables, we enabled agents to dynamically adapt their policies based on real-time constraint satisfaction levels. The integration of a gossiping protocol allowed agents to share local reward estimations across a stochastic network with one-bit communication, eliminating the need for direct state access and ensuring robust coordination. Our proposed contractive update rule for dual variables further enhanced scalability and feasibility without relying on exponentially growing memory buffers. We established theoretical guarantees of almost sure feasibility, as formalized in our main theorem, and validated our approach through  numerical experiments. In these experiments, a team of robots successfully patrolled multiple regions under realistic communication constraints and intermittent connectivity, showcasing the practical applicability of our framework.

\appendix \label{sec:appendix}
\renewcommand{\theequation}{\thesection.\arabic{equation}}
\setcounter{equation}{0}

\renewcommand{\thefigure}{\thesection.\arabic{figure}}
\setcounter{figure}{0}

\renewcommand{\thetable}{\thesection.\arabic{table}}
\setcounter{table}{0}
\section{Proof of Proposition \ref{prop_gradients}}
\label{app:proof_prop_gradients}
\begin{proof}(Proposition \ref{prop_gradients})
 The separable structure of our policy $\pi_\theta(A_t\mid S_t,\lambda,\ldots,\lambda) = \prod_{n=1}^N \pi_{\theta_n}(A_{tn}\mid S_{tn},\lambda)$ yields  $\nabla_{\theta_n}\log \pi_\theta(S_t,A_t) = \nabla_{\theta_n}\log \pi_{\theta_n}\left(A_{tn}\mid S_{tn}\right).$ 
 Thus, we obtain
 $\nabla_{\theta_n} \mathcal{L(\pi_\theta,\lambda)} = \mathbb E\left[L_nQ_{\pi_\theta}(S_t,A_t,\lambda)\right].$
 by substituting $L_n=\nabla_{\theta_n }\log \pi_{\theta_n}(A_{tn}\mid S_{tn})$  in the policy gradient \cite{sutton2000policy}, Next, we use the law of total probability  conditioning
on $S_{tn}$,$A_{tn}$, and write
$ \nabla_{\theta_n} \mathcal{L(\pi_\theta,\lambda)}$ 
$= \mathbb E_{S_{t}^n,A_{t}^n}\left[L_n\mathbb E_{(S_t,A_t)}\left[Q_{\pi_\theta}(S_t,A_t,\lambda)\mid S_{t}^n,A_{t}^n\right]\right],$
Then, we conclude the proof by noticing that the inner expectation coincides with $Q_{\pi_\theta}^n$ as defined in \eqref{eqn_q_others}.
\end{proof}
\section{Proof of Proposition \ref{prop_binomial}}
\label{app:proof_prop_binomial}
\begin{proof}(Proposition \ref{prop_binomial})
Consider two agents $J\leq d_G$ nodes apart. The waiting times $\left(t_1, \hdots, t_J\right)$ to cross the intermediate edges are i.i.d. geometric, so a journey has probability
$p\left(t_1, \hdots, t_J\right)= \prod_{j=1}^{J} p(1-p)^{t_j-1} = p^J(1-p)^{\sum_j t_j-J},$
where $p$ is the stochastic graph's communication probability. Since for any journey this depends only on the total travel time $i=\sum_j t_j$, the probability that $r(S_{\tau})$ reaches agent $n$ at time $i$ is $\binom{i-1}{J-1}p^J(1-p)^{i-J}=BN_{p,J}$.
Hence, the reward $\left(r(S_{\tau})\right)_m$ related to the $m$-th constraint at time $\tau$ will reach agent $n$ on time at time $(k+1)T_0-1$ or before  with probability
\begin{align}\mathbb P\left((e_\tau)_m=0\right)&=\mathbb P\left(\left(r(S_{\tau})=R_{\tau,(k+1)T_0-1}^n\right)_m\right) \label{eqn_gossip_error_def}\\
&\geq \mathbb P(BN(J,p)\leq (k+1)T_0-1-\tau)\nonumber\\
&\geq \mathbb P(BN(d_G,p)\leq (k+1)T_0-1-\tau),\nonumber
\end{align}
 where the first inequality holds since the error $e_\tau=r(S_{\tau})-R_{\tau,(k+1)T_0-1}^n$ is null if $r(S_{\tau})=0$, even if the message fails to reach agent $n$, and the second one is because the probability of a negative binomial reduces with the number of required successes. Finally, rewriting \eqref{eqn_gossip_error_def} as 
 \hspace{-10pt} \begin{align}\hspace{-8pt}\mathbb P\hspace{-1pt}\left(\hspace{-1pt}\left(e_\tau\right)_m=1\right)\leq\mathbb P\left(\hspace{-1pt}BN(d_G,p)\hspace{-1pt}>\hspace{-1pt}(k+1)T_0\hspace{-2pt}-\hspace{-2pt}1\hspace{-2pt}-\hspace{-2pt}\tau\right).\label{eqn_prob_gossip_err}
 \end{align}
 we obtain the desired result and conclude the proof.
\end{proof}

\section{Proof of Proposition \ref{prop_gossip_error}}\label{app:proof_prop_gossip_error}
\begin{proof}(Proposition \ref{prop_gossip_error})
We use the previous result to bound the expected error during a rollout. Since  the entries of $e_\tau$ in \eqref{eqn_gossip_error_def} can only take  values $0$ or $1$, it holds 
\begin{align}
&\mathbb E\left[\textstyle\sum_{t=k T_0}^{(k+1)T_0-1} \left(r(S_{\tau})-R_{\tau,(k+1)T_0-1}^n\right)_m\right]\\
&= \textstyle\sum_{\tau=k T_0}^{(k+1)T_0-1} \mathbb P\left(\left(e_\tau\right)_m=1\right)\\
&\leq  \textstyle\sum_{\tau =k T_0}^{(k+1)T_0-1} \mathbb P\left(BN(d_G,p)> (k+1)T_0-1-\tau\right)\nonumber\\
&=  \textstyle\sum_{l=0}^{T_0} \mathbb P\left(BN(d_G,p)> l\right)\leq   \textstyle\sum_{l=0}^{\infty} \mathbb P\left(BN(d_G,p)> l\right)\nonumber\\
&=\mathbb E\left[BN(d_G,p)\right]=\frac{d_G}{ p},\label{eqn_expectation_BN}
\end{align}
where we used \eqref{eqn_prob_gossip_err},  substituted $l=(k+1)T_0-1-\tau$, and added the tail to the sum so that it equals the expectation in \eqref{eqn_expectation_BN}, which is $d_G/p$ for the negative binomial.
%
To associate this expected value with the error between multipliers, we consider a particular agent $n$, take the difference between the global and local updates in \eqref{eqn_contractive_big_brother} and \eqref{eqn_stochastic_dual}, and bound the expectation as  
%
\begin{align}
    &\mathbb E\left[\|\lambda_{k+1}^{n}-\lambda_{k+1}\|_\infty \right]\label{eq:rolled_lambda_error}\\
    &\leq \frac{\eta}{T_0}  \mathbb E\left[ \|\textstyle\sum_{\tau=k T_0}^{(k+1) T_0-1}\left(r(S_{\tau})-R_{\tau,(k+1)T_0-1}^n\right)\|_\infty\right]\nonumber\\
    &+(1-\alpha) \mathbb E\left[\|\lambda_{k}^{n}-\lambda_{k}\|\right]\leq \frac{ \eta d_G}{T_0 p}+(1-\alpha)\mathbb E\left[\|\lambda_{k}^{n}-\lambda_{k}\|_\infty\right].\nonumber
\end{align}
Unrolling \eqref{eq:rolled_lambda_error} with $\lambda_0^n=\lambda_0$, we obtain
\begin{align}\nonumber
\mathbb E\left[\|\lambda_{k+1}^{n}-\lambda_{k+1}\|\right]&\leq \sum_{i=0}^k (1-\alpha)^i \frac{ \eta d_G}{T_0 p}
\leq \frac{ \eta d_G}{T_0 \alpha p},
\end{align}
which concludes the proof.
\end{proof}

\section{Proof of Proposition \ref{proposition:combined_lemma}}\label{app:combined_lemma}
The proof of Proposition  \ref{proposition:combined_lemma} relies on the following four Lemmas, which show that the multipliers follow a contractive evolution in a martingale sense  (Lemma \ref{lemma:almost_a_supermartingale}), allowing us to prove bound their expected running averages  by $\mathcal O(1/\sqrt{\alpha})$ (Lemma \ref{lemma_expected_mult_running_average}), and by a deterministic constant (Lemma \ref{lemma:positive_ideal_value}). This deterministic bound is a technicality to show that we can drop the expectation from Lemma \ref{lemma_expected_mult_running_average} and bound the stochastic running averages by $\mathcal O(1/\sqrt{\alpha})$ (Lemma \ref{lemma_expected_to_stochastic}) as desired. 
\begin{lemma}\label{lemma:almost_a_supermartingale}
  {Given  the stochastic communication network model \eqref{eqn_graph_model}, under Assumptions \ref{assumption_representation}--\ref{assumption_noforces}, and  with $\|c\|_\infty<1$, $\|c\|_1\leq N-1$, $\delta=\left(1-\|c\|_\infty\right) - M\left(\beta+\varepsilon_{T_0}\right)>0$ and $\alpha\geq\frac{\eta d_G}{p T_0}\frac{M L}{\delta}$, the multipliers satisfy the following inequality}
   \begin{align}
  {\mathbb E\left[\left\|\lambda_k\right\|^2 \mid \mathcal{F}_{k-1}\right] \leq
  (1-\alpha)^2\left\|\lambda_{k-1}\right\|^2+\eta^2 M}.
  \label{eq:almost_a_supermartingale}
  \end{align}
  \end{lemma}
\begin{proof}
As in the proof of Theorem \ref{theorem:feasibility} we denote the average reward during the k-th rollout as $\bar{r}_k^{T_0}=\frac{1}{T_0} \sum_{\tau=(k-1) T_0}^{k T_0-1} r\left(S_t\right)$ and rewrite  the   recursion \eqref{eqn_contractive_big_brother} as
\begin{align}\label{eqn_contractive_bb_rewritten}
& \lambda_{k+1}=\left[(1-\alpha) \lambda_k+\eta\left(c-\bar{r}_k^{T_0}\right)\right]_{+}.
\end{align}
Then, we use that projections reduce distances, together with $(c-\bar{r}_k^{T_0})_m^2<1$, to bound
\begin{align*}
&\mathbb E[ \|\lambda_{k+1}\|^2\hspace{-1pt}\mid\hspace{-1pt}\mathcal F_k]\hspace{-2pt}=\hspace{-2pt}\mathbb E\left[\hspace{-1pt}\|\hspace{-1pt}[(1-\alpha) \lambda_k+\eta(c-\bar{r}_k^{T_0})]_{+}\hspace{-1pt}\|^2\hspace{-2pt}\mid\hspace{-2pt}\mathcal F_k\hspace{-1pt}\right]\\
&\leq  \mathbb E\left[ \|(1-\alpha) \lambda_k+\eta(c-\bar{r}_k^{T_0})\|^2\mid \mathcal F_k\right]\\
&\leq  (1-\alpha)^2\left\| \lambda_k\right\|^2 +2(1-\alpha)\eta \lambda_k^T \mathbb E[c-\bar{r}_k^{T_0}\mid \mathcal F_k]
\\& \hspace{9pt}+\eta^2 \mathbb E\left[\|c-\bar{r}_k^{T_0}\|^2\mid \mathcal F_k\right]\\
&\leq  (1-\alpha)^2\left\| \lambda_k\right\|^2 +2(1-\alpha)\eta \lambda_k^T \mathbb E\left[c-\bar{r}_k^{T_0}\mid \mathcal F_k\right]+\eta^2 M.
\end{align*}
The expectation above must be taken with respect to the realizable policy $\pi_\theta[\lambda_k^1\ldots,\lambda_k^N]$ in \eqref{eqn_realizable_policy}, since we want to obtain results for stochastic trajectories following that policy. Thus  $\mathbb E\left[\bar{r}_k^{T_0}\mid \mathcal F_k\right]=V_R$ as defined in \eqref{eq:non-idealities-VR}, so that 
\begin{align}
\mathbb E\left[ \|\lambda_{k+1}\|^2\mid \mathcal F_k\right]& \leq   (1-\alpha)^2\left\| \lambda_k\right\|^2\label{eq:almost_supermartinagale_withVR} \\&+2(1-\alpha)\eta \lambda_k^T \left(c-V_R\right)+\eta^2 M,\nonumber
\end{align}
and by  \eqref{eqn_real_accute}, under the hypotheses of Lemmas \ref{lemma:positive_ideal_value} and \ref{lemma:error_gradient},  $\lambda_k^T \left(c-V_R\right)\leq 0$, which concludes the proof by removing the second term in the RHS of  \eqref{eq:almost_supermartinagale_withVR}. 
\end{proof}
\begin{lemma}\label{lemma_expected_mult_running_average}
  Under the hypothesis of Lemma \ref{lemma:almost_a_supermartingale} \begin{align}\label{eq:slln_bound}
  \limsup _{k \rightarrow \infty} \frac{1}{k} \sum_{i=1}^{k} \mathbb E\left[\left(\lambda_i\right)_m \right] \leq  \eta \sqrt{\frac{M}{\alpha}}.
  \end{align}
  \end{lemma}
\begin{proof} 
Using \eqref{eq:almost_a_supermartingale}   we obtain the inequality  $
\mathbb E[\|\lambda_k \|^2]$$\leq (1-\alpha)^2 \mathbb E\left[\|\lambda_{k-1}\|^2\right]+\eta^2M.$
Thus, by induction starting from $\lambda_0=0$ for convenience 
\begin{align}
 \mathbb E\left[\|\lambda_k \|^2\right]&\leq \textstyle\sum_{j=0}^{k-1}(1-\alpha)^{2j}\eta^2M\leq \eta^2 M \textstyle\sum_{j=0}^{\infty}(1-\alpha)^{2j}\nonumber\\
 &= \frac{\eta^2M}{(1-(1-\alpha)^2)}
 \leq  \eta^2M/\alpha.   \label{eq:bound_the_moment}
\end{align}
Applying Jensen's inequality and then  \eqref{eq:bound_the_moment}, we can bound the moment of $\lambda_{k+1}$ by
$
\left(\mathbb E\left[\|\lambda_{k+1}\|\right]\right)^2 \leq
\mathbb E\left[\| \lambda_{k+1} \|^2\right]\leq \eta^2M/\alpha,$
which yields the following bound for the expected value of each entry of  $\lambda_k$,   $\mathbb E\left[(\lambda_k)_m\right] \leq \mathbb E\left[\left\|\lambda_k\right\|\right] \leq \eta\sqrt{M/\alpha}$.
\end{proof}

In the next lemma, we prove that the multipliers are deterministically bounded by a constant $U$ of order $\mathcal O(1/\alpha)$, deriving this result directly from the contractive form of the stochastic update \eqref{eqn_contractive_big_brother}. Thus, it is immediate that the stochastic time average \eqref{eqn_bound_stochasstic_average} in Proposition \ref{proposition:combined_lemma} is bounded by $U$ too. Although more complex, the argument in the proof of Proposition \ref{proposition:combined_lemma} is meaningful because it yields a tighter bound of order $\mathcal O(1/\sqrt{\alpha})$ as in \eqref{eqn_bound_stochasstic_average}. Hence, the feasibility error \eqref{eqn_as_feasibility} in Theorem \ref{theorem:feasibility}, which results from multiplying the bound \eqref{eqn_bound_stochasstic_average} by $\alpha$ in \eqref{eq:constraint_satisfaction_mean}, can be controlled by the design parameter $\alpha$.

\begin{lemma}\label{lemma_deterministic_bound}
 Under the hypothesis of Lemma \ref{lemma:almost_a_supermartingale}, the multipliers $\lambda_k$ in \eqref{eqn_contractive_big_brother} are bounded a.s. by $U=\eta \sqrt{M}/\alpha$.
\end{lemma}
\begin{proof}  From the update \ref{eqn_contractive_big_brother} written as in \ref{eqn_contractive_bb_rewritten} , yields
\begin{align*}
&\left\|\lambda_{k+1}\right\|  \leq\left\|\lambda_k(1-\alpha)+\eta\left(c-\bar{r}_k^{T_0}\right)\right\| \\
& \leq(1-\alpha)\left\|\lambda_k\right\|+\eta\left\|c-\bar{r}_k^{T_0}\right\|  \leq(1-\alpha)\left\|\lambda_k\right\|+\eta \sqrt{M}\\
& \leq(1-\alpha)^k\left\|\lambda_0\right\|+ \textstyle\sum_{j=0}^k \eta \sqrt{M}(1-\alpha)^j \\
& \leq \eta \sqrt{M}  \textstyle\sum_{j=0}^{\infty}(1-\alpha)^j  =\eta \sqrt{M}/\alpha,
\end{align*}
where we assumed $\lambda_0=0$ for convenience.
\end{proof}

\begin{lemma}\label{lemma_expected_to_stochastic}
  Assume that there exists a constant $C >0$ such that $\limsup _{k \rightarrow \infty} \frac{1}{k} \sum_{i=1}^{k} \mathbb E\left[\left(\lambda_i\right)_m \right] \leq C$.
  Then the following bound also holds a.s. for the average  stochastic multipliers, i.e., $\limsup_{k \rightarrow \infty} \frac{1}{k} \sum_{i=1}^{k} \left(\lambda_i\right)_m \leq C.$
\end{lemma}

\begin{proof} 
Define   $\ell_{k+1} :=\lambda_{k+1}-\mathbb E\left[\lambda_{k+1} \mid \mathcal F_k\right]$. We start by proving that these  $\ell_k$ satisfy the SLLN. By definition, these variables are zero-mean conditioned on $\mathcal F_k$, and deterministically bounded by $2U$ according to Lemma \ref{lemma_deterministic_bound}, but not i.i.d. because $\lambda_k$ accumulates rewards.  To prove
 $ \mu_{k}:=\frac{1}{k+1} \sum_{i=0}^{k} \ell_i \xrightarrow{a. s.} 0,
  $
 consider the partial sums   
     $U_{k+1}=\sum_{i=1}^{k+1} \ell_i=\ell_{k+1}+U_k $. Since  $\ell_{k+1}$ are conditionally zero-mean,  $U_k$ is a martingale, i.e.,
    $
      \mathbb E\left[U_{k+1}\mid \mathcal F_k \right]=\mathbb E\left[\ell_{k+1}\mid \mathcal F_k \right]+U_k=U_k.
     $
     
     To use the proof of the SLLN in \cite[p.456]{Gallager}, we need a bound for the second moment of $U_k$  that increases with order $k$ at most. Without the i.i.d. assumption, we write
     \begin{align*}
    \mathbb E\left[\left\|U_{k+1}\right\|^2\mid \mathcal F_k\right] & =\left\|U_k\right\|^2+2 U_k^{\top} \mathbb E\left[\ell_{k+1} \mid \mathcal F_k\right]\\
    &+\mathbb E\left[\|\ell_{k+1}\|^2\mid \mathcal F_k\right]  \leq\left\|U_k\right\|^2+U^2.
    \end{align*}
    Conditioning to $\mathcal F_{k-1}$ instead, we bound
    \begin{align*}
     &\mathbb  E\left[|| U_{k+1} \|^2\mid \mathcal F_{k-1}\right]=\mathbb  E\left[\mathbb E\left[|| U_{k+1} \|^2\mid \mathcal F_{k}\right]\mid\mathcal F_{k-1}\right]\\ 
    &\leq \mathbb  E\left[|| U_k \|^2+U^2 \mid \mathcal F_{k-1}\right]  \leq \mathbb  E\left[\left\|U_k\right\|^2 \mid {\mathcal F_{k-1}}\right]+U^2  \\
    & \leq\left\|U_{k-1}\right\|^2+U^2+U^2. 
    \end{align*}
    If we repeat this process recursively, reducing the time index of the filtration, we arrive at $\mathbb E\left[|| U_{k+1} \|^2\mid \mathcal F_{0}\right]=\mathbb E\left[|| U_{k+1} \|^2\right]$, obtaining  a bound for the unconditioned moment of $U_{k+1}$:   
    \begin{align*} \mathbb E\left\|U_{k+1}\right\|^2 &\leq \mathbb E\left\|U_1\right\|^2+k U^2 = \mathbb  E\left\|\lambda_0\right\|^2+(k+1) U^2  \\&\leq U^2+ k U^2  \leq(k+1) U^2.
    \end{align*}
    Henceforth, we substitute $X_i$ and  $S_n$  for $\ell_i$ and $U_k$, respectively, and follow the proof in \cite[p.456]{Gallager}  to conclude
 \begin{align*}
    \mu_{k}&:=\frac{1}{k+1} \sum_{i=0}^{k} \ell_i=\frac{1}{k+1} \sum_{i=0}^{k}\left(\lambda_i-\mathbb E \left[\lambda_i\mid \mathcal F_{i-1}\right)\right] \xrightarrow{a. s.} 0.
    \end{align*}
    Next, we want to get rid of the filtrations and compare the stochastic running average to the average of unconditional expectations, i.e.,
    $\frac{1}{k} \sum_{i=1}^{k}\left(\lambda_i-\mathbb E \left[\lambda_i\right] \right)\xrightarrow{a. s.} 0.
  $
  To that end, consider $\bar{\lambda}_{k+1}^{(k)}=\mathbb E\left[\lambda_{k+1}\mid  F_k\right]=\mathbb E\left[\lambda_{k+1}\mid \lambda_k\right]$,
  which is a random variable that is a function of $\lambda_k$ for $k>1$,
  and with $\mathbb  E\left[\lambda_k \mid \mathcal F_{k-1}\right]=\mathbb  E\left[\lambda_i\right]$ if $k \leq 1$.
Thus, we can use a recursive argument by redefining the conditionally zero-mean $$\ell_{k+1}\hspace{-1pt}=\hspace{-1pt}\bar\lambda_{k+1}^{(k)}-\mathbb E\hspace{-2pt}\left[\hspace{-1pt}\bar\lambda_{k+1}^{(k)}\hspace{-3pt}\mid\hspace{-3pt}\mathcal F_{k-1}\right]\hspace{-3pt}=\hspace{-1pt}\mathbb E\hspace{-1pt}\left[\lambda_{k+1}\hspace{-2pt}\mid\hspace{-2pt}\mathcal F_k\right]-\mathbb E\hspace{-1pt}\left[\lambda_{k+1}\hspace{-3pt}\mid\hspace{-3pt}\mathcal F_{k-1}\right],$$ which are also bounded by $U$, hence we can apply the SLLN again to the corresponding time averages
  \begin{align} \nonumber\mu_{k}\hspace{-1pt}&=\hspace{-1pt}\frac{1}{k}\hspace{-1pt}\sum_{i=1}^{k}\hspace{-1pt}\ell_i\hspace{-1pt}=\hspace{-1pt}\frac{1}{k}\hspace{-1pt}\sum_{i=1}^{k}\hspace{-1pt}\left(\mathbb E\left[\lambda_{i}\hspace{-1pt}\mid\hspace{-1pt}\mathcal F_{i-1}\hspace{-1pt}\right]\hspace{-1pt}-\hspace{-1pt}\mathbb E\hspace{-1pt}\left[\lambda_{i}\hspace{-1pt}\mid\hspace{-1pt}\mathcal F_{i-2}\right] \right)\hspace{-1pt}\xrightarrow[k\hspace{-1pt}\rightarrow\hspace{-1pt}\infty]{a.s.}\hspace{-1pt}0.
  \end{align}
  Then we can move backwards on time by repeating this argument for previous $i-n+1$ and $i-n$, i.e.,
  \begin{align}
  \mu_{k}=\frac{1}{k} \sum_{i=1}^{k}\left(\mathbb E\left[\lambda_{i} \mid \mathcal F_{i-n+1}\right]-\mathbb E\left[\lambda_{i} \mid \mathcal F_{i-n}\right] \right)\xrightarrow[k \rightarrow \infty]{a.s.} 0,\nonumber
  \end{align}
   and  the conditioning can be removed when we reach $n\geq k$, resulting in
  $
   \mathbb E\left[\lambda_i\mid \mathcal F_{i-k}\right]=\mathbb E\left[\lambda_i\right]$, for all $i\leq k$, which implies that the stochastic and expected running averages coincide. Hence, our result follows.
  \end{proof}
Finally, put lemmas together to prove the proposition.
\begin{proof}(Proposition \ref{proposition:combined_lemma})
    Substitute $C=\eta\sqrt{{M}/{\alpha}}$ from Lemma \ref{lemma_expected_mult_running_average} in Lemma \ref{lemma_expected_to_stochastic}. 
\end{proof}


\printbibliography
\begin{IEEEbiography}
[{\includegraphics[width=1in,height=1.25in,clip,keepaspectratio]{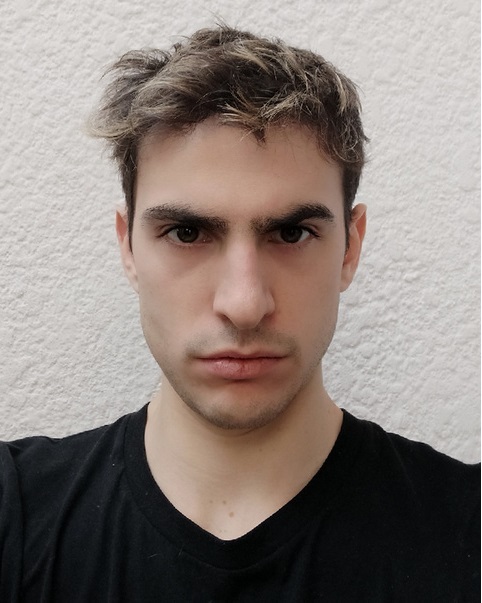}}]
{Leopoldo Agorio} received the B.Sc. degree in Physics and the B.Sc. and M.Sc. degrees in Electrical Engineering from Universidad de la República (UdelaR), Montevideo, Uruguay, in 2018, 2019, and 2022, respectively. He is currently pursuing the Ph.D. degree at the same institution, where he serves as teaching and research assistant at the Department of Electrical Engineering. His current research interests include mobile robotics, multi-agent reinforcement learning, and quantum computing. He also has industry experience in robotics

and computer vision for quality control.
\end{IEEEbiography}

\begin{IEEEbiography}[{\includegraphics[width=1in,height=1.25in,clip,keepaspectratio]{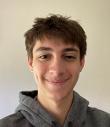}}]
{Sean Van Alen} received the B.Sc. degree in Electrical and Computer Engineering from the University of Pittsburgh (Pitt) in May 2025. Sean received the 2024 Scholarship for Undergraduate Research from the department of Biomedical Engineering at Pitt, under the supervision of Dr. Bazerque.
\end{IEEEbiography}

\begin{IEEEbiography}[{\includegraphics[width=1in,height=1.25in,clip,keepaspectratio]{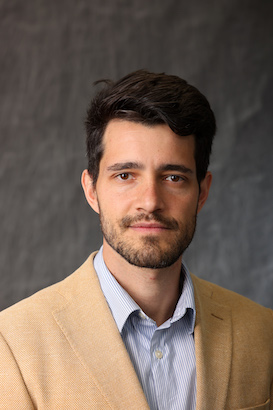}}]
{Santiago Paternain }
received the B.Sc. degree in electrical engineering from Universidad de la República, Montevideo, Uruguay in 2012, the M.Sc. in Statistics from the Wharton School in 2018 and the Ph.D. in Electrical and Systems Engineering from the Department of Electrical and Systems Engineering, University of Pennsylvania in 2018. He is currently an Assistant Professor in the Department of Electrical Computer and Systems Engineering at Rensselaer Polytechnic Institute. Prior to joining Rensselaer, Dr. Paternain was a postdoctoral Researcher at the University of  Pennsylvania. His research interests lie at the intersection of machine learning and control of dynamical systems. Dr. Paternain was the recipient of the 2017 CDC Best Student Paper Award and the 2019 Joseph and Rosaline Wolfe Best Doctoral Dissertation Award from the Electrical and Systems Engineering Department at the University of Pennsylvania. 
\end{IEEEbiography}

\begin{IEEEbiography}[{\includegraphics[width=1in,height=1.25in,clip,keepaspectratio]{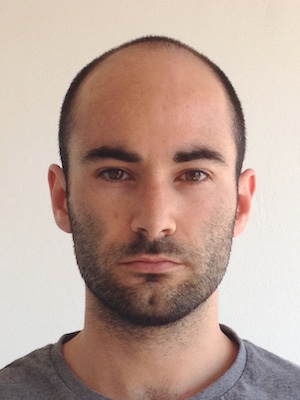}}]
{Miguel Calvo-Fullana} received the B.Sc. degree in Electrical Engineering from the Universitat de les Illes Balears (UIB), Spain, in 2010, and the M.Sc. and Ph.D. degrees in Electrical Engineering from the Universitat Politècnica de Catalunya (UPC), Spain, in 2013 and 2017, respectively. He joined Universitat Pompeu Fabra (UPF), Barcelona, Spain, in 2023, where he is a Ramón y Cajal Fellow. Prior to joining UPF, he held postdoctoral appointments at the University of Pennsylvania and the Massachusetts Institute of Technology. During his Ph.D. studies, he was a research assistant at the Centre Tecnològic de Telecomunicacions de Catalunya (CTTC). His research interests lie broadly in learning and optimization for autonomous systems, with a particular emphasis on multi-agent systems, wireless communications, and network connectivity. He is the recipient of Best Paper Awards at ICC 2015, GlobalSIP 2015, and ICASSP 2020.
\end{IEEEbiography}

\begin{IEEEbiography}[{\includegraphics[width=1in,height=1.25in,clip,keepaspectratio]{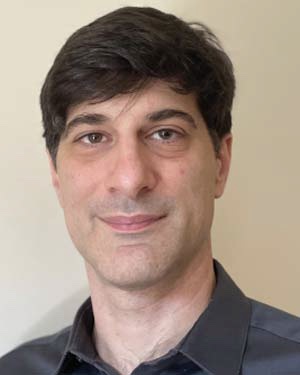}}]
{Juan Andr\'es Bazerque} received the B.Sc. degree in Electrical Engineering from the Universidad de la República (UdelaR), Montevideo, Uruguay, in 2003, and the M.Sc. and Ph.D. degrees in Electrical Engineering from the Department of Electrical and Computer Engineering at the University of Minnesota, Minneapolis, MN, USA, in 2010 and 2013, respectively. Since 2015, he has been an Assistant Professor with the Department of Electrical Engineering at UdelaR. In 2026, he joined the Department of Electrical and Computer Engineering at Carnegie Mellon University, Pittsburgh, PA, USA, as a Special Research Professor. His current research interests include stochastic optimization and networked systems, with a focus on reinforcement learning, graph signal processing, and quantum computing. Dr. Bazerque is the recipient of the University of Minnesota's Master's Thesis Award for 2009–2010 and a co-recipient of the Best Paper Award at the 2nd International Conference on Cognitive Radio Oriented Wireless Networks and Communications in 2007.
\end{IEEEbiography}
\end{document}